\documentclass[a4paper,11pt,oneside]{article}

\usepackage[margin=1in]{geometry}
\usepackage{physics}
\usepackage{enumerate}
\usepackage{empheq}
\usepackage{booktabs}
\usepackage{amsmath,amssymb,amsbsy,amstext, amsthm, simplewick}
\usepackage{hyperref}
\usepackage{graphicx}
\usepackage{amsfonts}
\usepackage{ulem}
\usepackage{color}
\usepackage[svgnames,dvipsnames,x11names]{xcolor}
\usepackage{wasysym}
\usepackage{bbm}
\usepackage{bbold}
\usepackage{bm}
\usepackage{relsize}
\usepackage{cite}

\usepackage{colortbl}
\definecolor{summersky}{cmyk}{0.71,0.33,0,0.5}
\definecolor{flamingo}{cmyk}{0,0.51,0.71,0.5}
\definecolor{rp}{cmyk}{0.2, 1, 0.6, 0}
\definecolor{pacificblue}{cmyk}{0.95,0.3,0, 0.5}
\definecolor{gray60}{cmyk}{0.4,0.4,0,0.8}

\hypersetup{
    pdftoolbar=true,        
    pdfmenubar=true,        
    pdffitwindow=true,     
    pdfstartview={FitH},    
    pdfnewwindow=true,      
    colorlinks=true,       
    linkcolor=pacificblue,          
    citecolor=flamingo,        
    filecolor=magenta,      
    urlcolor=pacificblue           
}

\usepackage[framemethod=default]{mdframed}
\newmdenv[skipabove=7pt,
skipbelow=7pt,
rightline=false,
leftline=false,
topline=false,
bottomline=false,
backgroundcolor=pacificblue!10,
linecolor=gray,
innerleftmargin=5pt,
innerrightmargin=5pt,
innertopmargin=2pt,
innerbottommargin=10pt,
leftmargin=0cm,
rightmargin=0cm,
linewidth=4pt]{eBox}

\newcommand{\ex}[1]{\langle #1 \rangle}
\newcommand{\be}{\begin{eqnarray} }
\newcommand{\ee}{\end{eqnarray} }
\newcommand{\bs}{\begin{split} }
\newcommand{\es}{\end{split} }

\renewcommand{\L}{\mathcal{L}}

\newcommand{\e}{\epsilon}

\newcommand{\nn}{\nonumber}

\renewcommand{\(}{\left(}
\renewcommand{\)}{\right)}
\renewcommand{\[}{\left[}
\renewcommand{\]}{\right]}

\begin{document}


\begin{titlepage}
\baselineskip=15.5pt 
\thispagestyle{empty}

\begin{center}
{\fontsize{15}{22}\bf From Amplitudes to Contact Cosmological Correlators}\\
\end{center}

\vskip 18pt
\begin{center}
\noindent
{\fontsize{12}{18}\selectfont James Bonifacio\footnote{\tt jb2389@cam.ac.uk}, Enrico Pajer\footnote{\tt
			enrico.pajer@gmail.com} and Dong-Gang Wang\footnote{\tt dgw36@cam.ac.uk}}
\end{center}

\begin{center}
  \vskip 8pt
\textit{Department of Applied Mathematics and Theoretical Physics, University of Cambridge, Wilberforce Road, Cambridge, CB3 0WA, UK}
\end{center}

\vspace{1.4cm}

\noindent Our understanding of quantum correlators in cosmological spacetimes, including those that we can observe in cosmological surveys, has improved qualitatively in the past few years. Now we know many constraints that these objects must satisfy as consequences of general physical principles, such as symmetries, unitarity and locality. Using this new understanding, we derive the most general scalar four-point correlator, i.e., the trispectrum, \textit{to all orders in derivatives} for manifestly local \textit{contact} interactions. To obtain this result we use techniques from commutative algebra to write down all possible scalar four-particle amplitudes without assuming invariance under Lorentz boosts. We then input these amplitudes into a \textit{contact reconstruction formula} that generates a contact cosmological correlator in de Sitter spacetime from a contact scalar or graviton amplitude. We also show how the same procedure can be used to derive higher-point contact cosmological correlators. Our results further extend the reach of the boostless cosmological bootstrap and build a new connection between flat and curved spacetime physics.


\end{titlepage}

\setcounter{page}{2} 

\newpage
\setcounter{tocdepth}{2}
\tableofcontents

\newpage

\section{Introduction}

At the heart of our understanding of gravity sits the tenet that in a small enough neighbourhood of a generic spacetime point we should recover flat-spacetime physics. This gives us an entry point to define, at least perturbatively, quantum field theory around curved spacetime and a perturbative regime of quantum gravity described by the interacting theory of massless spin-2 particles. The challenge is to understand the global behaviour of the system from this local description. For example, we want to understand how the very uneventful experience of a scientist in a small lab falling into a large black hole can be compatible with and/or complementary to the description given by a far away observer. Furthermore, we expect that the flat-spacetime notion of \textit{consistent} (quantum field) theories is very different from the same notion in curved spacetimes. For example, in Minkowski space we know that there is a consistent effective field theory (EFT) of interacting massless spin-3/2 particles, namely supergravity with linearly realised supersymmetry. Conversely, such a theory does not exist in de Sitter space, where invariance under the action of supercharges is incompatible with a positive cosmological constant. As another example, the interaction of relativistic massless spin-2 particles in Minkowski space cannot break Lorentz boosts, neither explicitly nor spontaneously \cite{PSS}. Conversely, such a breaking of boosts is ubiquitous in cosmological models of dark energy and inflation, a fact which is particularly transparent in the EFT approach \cite{Creminelli:2006xe,EFTofI,Creminelli:2008wc,Gubitosi:2012hu}. Intuitively, to break boosts we need to fill spacetime with a medium, which, in the presence of dynamical gravity, curves spacetime and makes the Minkowski solution inconsistent. As a final example, while string theory constitutes a UV-complete theory for the scattering of gravitons in Minkowski space, there are doubts whether (stable, eternal) de Sitter spacetime can arise as a consistent solution of string theory or of any other quantum theory of gravity \cite{Danielsson:2018ztv}.

As the above discussion highlights, the connection between flat and curved spacetime physics, which is so important for our description of cosmology, is incredibly rich. Hence, any relations that we can find between the two setups can be useful. In the cosmological context, one such relation is given by the observation that, for accelerated FLRW cosmologies with a Bunch--Davies initial condition, there is a limit of cosmological correlators (or, equivalently, of the wavefunction coefficients) that contains Minkowski scattering amplitudes \cite{Maldacena:2011nz,Raju:2012zr} (see also Refs.~\cite{Arkani-Hamed:2018bjr,Benincasa:2018ssx}, and Ref.~\cite{COT} for an explicit derivation including overall factors). This relation arises in the so-called vanishing total-energy limit where momenta are analytically continued to the complex plane in such a way that only high-energy interactions of the fields contribute, for which the expansion of spacetime is a negligible correction. This limit already reveals some surprises. The amplitude appearing on the residue of the total-energy pole is not necessarily consistent in global Minkowski spacetime. One example of this is the theory of a superfluid coupled to gravity. This system does not admit Minkowski solutions and indeed the associated amplitudes do not factorise correctly \cite{PSS}. Another example is a canonical scalar field coupled to gravity. The cubic amplitude contained in the three-point function computed in Ref.~\cite{Maldacena:2002vr} contains inverse powers of derivatives and therefore cannot arise in any local theory in Minkowski space \cite{BBBB}. 

The vanishing total-energy limit tells us that if we throw away all terms in the correlators that know about the expansion of the universe (and hence do not conserve energy), then what is left is the flat-space amplitude. Here we want to ask about the opposite relation: \textit{given an amplitude, how can we write down a consistent corresponding correlator?} In this work, we make a small step towards answering this question. In particular, for massless scalars and gravitons in de Sitter space, we provide a contact reconstruction formula that takes a tree-level contact amplitude and outputs a contact wavefunction coefficient. These wavefunction coefficients are very closely related to the cosmological correlators that are constrained by observations. This formula does not assume invariance under de Sitter boosts and can therefore be used for all manifestly local interactions appearing in general models of inflation, including those in the EFT of inflation. At the technical level, the contact reconstruction formula can be thought of as \textit{a} solution of the manifestly local test, which is a condition that all de Sitter wavefunction coefficients of massless particles, including scalars and gravitons, must satisfy if they originate from interactions with only positive powers of derivatives \cite{Jazayeri:2021fvk}. The question of how to go from amplitudes to wavefunction coefficients was previously investigated using cosmological polytopes in Ref.~\cite{Benincasa:2018ssx}, building on Ref.~\cite{Arkani-Hamed:2017fdk}. While our motivations are similar, the results in this paper differ in a few aspects. First, Ref.~\cite{Benincasa:2018ssx} focussed on a class of toy models, namely scalar theories with only polynomial interactions, while here we consider arbitrary derivative interactions. Second, the explicit construction in Ref.~\cite{Benincasa:2018ssx} was given at the level of the \textit{integrand} of the wavefunction coefficients, which can be thought of as a Minkowski wavefunction (also recently discussed in Ref.~\cite{Baumann:2021fxj}), while here we give the result for the de Sitter wavefunction, \textit{after} computing the relevant time integral. Lastly, Ref.~\cite{Benincasa:2018ssx} was able to reconstruct all tree-level wavefunction coefficients from amplitudes, while in this work we only consider contact contributions.  

For phenomenological interest, correlators from inflation beyond the two-point function are the major targets of cosmological observations on primordial non-Gaussianity \cite{Meerburg:2019qqi}. It is therefore important to prepare a complete set of theoretical predictions that are allowed by fundamental physical principles. From this point of view, the bootstrap approach to cosmological correlators provides powerful tools to directly derive testable predictions without lengthy explicit computations for large classes of inflationary models \cite{CosmoBootstrap1,CosmoBootstrap2,Green:2020ebl, CosmoBootstrap3,PSS, BBBB,Stefanyszyn:2020kay,Jazayeri:2021fvk,Baumann:2021fxj,Gomez:2021qfd}. In this paper, as an application of our contact reconstruction formula, we bootstrap all possible manifestly local contact scalar trispectra at tree level to all orders in derivatives. This extends the results of Ref.~\cite{BBBB,Jazayeri:2021fvk} where all tree-level scalar bispectra were derived. These trispectra are precisely those that can arise in a generic theory of single-clock inflation, and their coefficients are related to the arbitrary coupling constants in the EFT at low energies. Some known results in the literature are reproduced, while new ones are derived as well.
Careful considerations of naturalness and non-linearly realised symmetries constrain the relative size of these trispectra, e.g., along the lines of Refs.~\cite{Senatore:2010jy,Bartolo:2010di,Smith:2015uia}, however we will not discuss this in the present work. As a technical note, we should mention that we work at the level of the IR finite part of the wavefunction coefficients, which are forced to be real by unitarity in the form of the cosmological optical theorem \cite{COT,Cespedes:2020xqq,Melville:2021lst,single,Baumann:2021fxj}---see also Refs.~\cite{Aharony:2016dwx,Meltzer:2019nbs,Meltzer:2020qbr} for the anti de Sitter (AdS) side of this story and Refs.~\cite{Sleight:2019mgd,Sleight:2019hfp,Sleight:2020obc} for connecting results in AdS space to dS space. 

The rest of the paper is organised as follows: in Section \ref{sec:2} we review our setup and notation and discuss the general rules that we use to bootstrap correlators. In Section \ref{sec:WFfromAmp} we derive our contact reconstruction formula, which given a contact $  n $-particle amplitude for massless scalars and gravitons generates a corresponding $  n $-point correlator or wavefunction coefficient in de Sitter space. As an illustration we show how this generates the scalar bispectrum to all orders in derivatives. In Section \ref{sec:trispectrum} we derive the scalar four-point wavefunction (equivalently, the trispectrum) for contact interactions to all orders in derivatives. We obtain this result by first constructing a basis for the polynomial ring of four-particle boost-breaking contact amplitudes and then using this in the contact reconstruction formula. In Section \ref{sec:5} we show how our methods can be extended to contact higher-point functions.
Finally, we conclude in Section \ref{sec:6}.

\paragraph{Notation and conventions:} We denote the elementary symmetric polynomials in the three variables $k_1, k_2, k_3$ with blackboard font
\begin{align}\label{1}
k_{T}= \mathbb{e}_{1}&\equiv k_{1}+k_{2}+k_{3}\,, \\
\mathbb{e}_{2}&\equiv k_{1}k_{2}+k_{1}k_{3}+k_{2}k_{3}\,, \\
\mathbb{e}_{3}&\equiv k_{1} k_{2} k_{3}\,,
\end{align}
while the elementary symmetric polynomials in the four variables $k_1, \dots, k_4$ are  denoted by
\begin{align}
k_{T}=e_{1}&\equiv k_{1}+k_{2}+k_{3}+k_{4}\,, \\
e_{2}&\equiv k_{1}k_{2}+k_{1}k_{3}+k_{2}k_{3}+k_{1}k_{4}+k_{2}k_{4}+k_{3}k_{4}\,, \label{eq:e2} \\
e_{3}&\equiv k_{1} k_{2} k_{3}+k_{1} k_{2} k_{4}+k_{1} k_{3} k_{4}+ k_{2} k_{3}k_{4}\,,\\
e_{4}&\equiv k_{1}k_{2}k_{3}k_{4}\,, \label{3esp}
\end{align}
where the meaning of $k_T$ should be clear from context.
We also define spatial Mandelstam-like variables as
\be \label{eq:Mandelstam}
s_{ij} \equiv (\vec{k}_i + \vec{k}_j)^2.
\ee
We denote correlation functions as
\begin{align}
\ex{\phi(\vec k_{1})\phi(\vec k_{2}) \dots \phi(\vec k_{n})}=(2\pi)^{3}\delta \! \left(  \vec k_{1}+\dots +\vec k_n\right)   B_{n} (\vec k_1, \dots, \vec k_n) \,.
\end{align}
We use a condensed notation for momentum integrals,
\be
\int_{\vec{k}} \equiv \int \frac{d^3 \vec{k}}{(2\pi)^3}.
\ee

 
\section{The rules of the game}\label{sec:2}

In this section, after a lightning introduction to the wavefunction of the universe and the Schr\"odinger picture approach to quantum field theory in de Sitter space, we review the main results that we will use in the rest of the paper, namely the bootstrap rules for boostless contact interactions in de Sitter space \cite{BBBB} and the manifestly local test recently derived in Ref.~\cite{Jazayeri:2021fvk}. Taking the boundary perspective of the bootstrap approach, here we mainly focus on
cosmological correlators evaluated in the asymptotic future, at the so-called (future, spacelike, conformal) boundary of de Sitter space. 

\subsection{The wavefunction of the universe}
Let us begin by introducing the main object of interest in this paper---the wavefunction of the universe $\Psi$ at the  late-time boundary of de Sitter space. For pedagogical discussions of the Schr\"odinger picture approach to QFT in de Sitter space, see, e.g., Refs.~\cite{Guven:1987bx,Maldacena:2002vr,WFCtoCorrelators1,WFCtoCorrelators2,COT}.
For a theory with a scalar field $\phi$, a general late-time wavefunction can be expressed as
\begin{equation}
\Psi [\phi({\vec k})] = \exp \[ -\sum_{n=2}^\infty \frac{1}{n!}
\int_{\vec k_1, \dots, \vec k_n} \psi_n(\vec k_1, \dots, \vec k_n) \,(2\pi)^{3}\delta \!\left(  \vec k_{1}+\dots +\vec k_n\right)\,  
\phi(\vec k_{1}) \dots \phi(\vec k_{n})  \] ,
\end{equation}
where $\psi_n(\vec k_1, \dots, \vec k_n) $ are the wavefunction coefficients and the presence of the momentum-conserving delta function is enforced by invariance under spatial translations.\footnote{In some conventions $\psi_n$ is defined with a delta function and a prime indicates that it has been stripped away. To avoid cluttering our notation, here we do not employ that convention.} The equal-time $n$-point correlation functions of the $\phi$ field can be computed from the wavefunction of the universe as 
\be
\ex{ \phi(\vec k_{1})\phi(\vec k_{2}) \dots \phi(\vec k_{n}) }=\frac{\mathlarger\int{\mathcal D}\phi \,\phi(\vec k_{1})\phi(\vec k_{2}) \dots \phi(\vec k_{n}) |\Psi [\phi ]|^2}{\mathlarger\int{\mathcal D}\phi \, |\Psi [\phi ]|^2}\,.
\ee
Therefore, by performing this integral over ``boundary'' fluctuations $ D\phi  $, which is possible at least perturbatively, one can compute the cosmological correlators directly from the wavefunction coefficients. For the power spectrum and bispectrum, these relations at tree level are 
\begin{align}
P(k)&=\frac{1}{2 \Re \psi_2(k)}\,, \\
 B_3(\vec k_1, \vec k_2, \vec k_3) &= -2\frac{\Re \psi_{3}(k_{1},k_{2},k_{3})}{\prod_{a=1}^{3}2\Re \psi_{2}(k_{a})}, \label{eq:WFCTo3Pt}
\end{align}
where $k_a \equiv|\vec{k}_a |$.
For four-point and higher-point correlation functions, the tree-level relations are more complicated. However, if we focus on {\it contact} contributions, namely correlators that are linear in a single coupling constant, they are given by\footnote{This formula is valid only for interactions with an even number of spatial derivatives. Conversely, interactions with an odd number of spatial derivatives pick up the imaginary part of $  \psi_{n} $ instead. Our focus here is scalar and graviton correlators in theories with parity-even interactions and therefore we can safely use this expression.}
\begin{align} \label{Bnpsin}
B_{n} (\vec k_1, \dots , \vec k_n) =-2\frac{\Re \psi_{n}(\vec k_1, \dots, \vec k_n)}{\prod_{a=1}^{n}2\Re \psi_{2}(k_{a})} .
\end{align}
Note that there can also be {\it exchange} contributions proportional to the product of lower-point contact terms. These can be bootstrapped from lower-point correlators and amplitudes---see Refs.~\cite{Jazayeri:2021fvk, Baumann:2021fxj,Gomez:2021qfd} for recent progress.
In this paper, we do not consider those contributions and focus exclusively on correlators/wavefunction coefficients from $n$-point contact interactions, which are simply related as in Eq.~\eqref{Bnpsin}. 

\subsection{Bootstrap rules}

Fundamental principles such as unitarity, symmetries, and locality severely constrain the form of wavefunction coefficients. Here we focus on contact $  n $-point correlators for massless scalars in a de Sitter background, assuming a Bunch--Davies vacuum \cite{Bunch:1978yq} and interactions that are invariant under rotations, translations and dilations, but not necessarily de Sitter boosts, which are broken by almost all inflationary models. This leads us to the following bootstrap rules \cite{BBBB}:
\begin{itemize}
\item \textbf{Tree-level calculations in (quasi) de Sitter space.} This implies that the $n$-point wavefunction coefficients are rational functions of the rotationally invariant contractions of the momenta $\vec k_a$ and of the energies $k_a\equiv \vert \vec k_a \vert$:
\begin{align}
\psi_n(\vec k_1, \dots, \vec k_n) \sim \frac{\text{Poly}_\alpha(\vec{k}_{a}\cdot \vec{k}_{b}, k_c)}{\text{Poly}_\beta(\vec{k}_{a}\cdot \vec{k}_{b}, k_c)}\, ,\label{poly}
\end{align}
where $\text{Poly}_\alpha$ denotes a homogeneous polynomial of homogeneity degree $\alpha$ under rescaling $\vec k_a$. For interactions with few derivatives, a logarithmic term may also appear and we will treat these cases separately in this paper.
\item \textbf{Scale invariance.} For massless fields in de Sitter space, scale invariance enforces that the wavefunction coefficients have an overall momentum scaling of $k^3$. Therefore, the rational function in Eq.~\eqref{poly} can be further reduced to
\begin{align}
\psi_n(\vec k_1, \dots, \vec k_n) \sim \frac{\text{Poly}_{\beta+3}(\vec{k}_{a}\cdot \vec{k}_{b}, k_c)}{\text{Poly}_\beta(\vec{k}_{a}\cdot \vec{k}_{b}, k_c)}\, .
\end{align}
\item \textbf{Bose symmetry.} For interactions with a single scalar field, Bose symmetry implies that the $n$-point function has to be symmetric under any permutation $  \vec{k}_{i} \leftrightarrow \vec{k}_{j} $.
For the bispectrum this means that it can be written in terms of elementary symmetric polynomials. In Section \ref{sec:trispectrum} we  work out the invariant polynomials needed for the trispectrum.

\item \textbf{Bunch--Davies vacuum and locality.} This implies that  for contact diagrams the only singularities that can appear occur when the total energy vanishes,\footnote{Here and in the following we assume that all fields have the same speed of sound, which we can set to unity. The case of different speeds is easily derived with an appropriate rescaling, as explained at the end of Section~\ref{ssec:WFCfromAmps}.} $ k_{T}\to 0 $. Thus we can write the rational function as a sum of terms with $k_T$ poles of various orders,
\begin{align}
\psi_{n} = \sum_{m=0}^p \frac{\text{Poly}_{3+p-m}({\vec{k}_{a}\cdot \vec{k}_{b}, k_c})}{ k_{T}^{p-m}} ,
\end{align}
where for manifestly local interactions the residues of the various $k_T$ poles are related by the manifestly local test to be discussed momentarily. 

\item \textbf{The amplitude limit.} The residue of the leading $k_T$ pole is proportional to the flat-space amplitude \cite{Maldacena:2011nz,Raju:2012zr}. This is the starting point for our reconstruction of wavefunction coefficients from amplitudes. We shall elaborate on this rule in the next section.

\end{itemize}

As we will see, these bootstrap rules enforce strong constraints on the form of the wavefunction coefficients. However, various free coefficients remain. 
For instance, although the leading $k_T$-pole term can be fixed by a corresponding scattering amplitude in flat space,  the residues of the subleading $k_T$ poles are still undetermined by these rules.
With the additional input of the manifestly local test, we can fix precisely these subleading total-energy terms.

\subsection{The manifestly local test}
We now review an additional constraint that applies to wavefunction coefficients for massless scalars with manifestly local interactions, i.e., interactions that are polynomials of propagating fields and their derivatives \textit{at the same spacetime point}. This restriction excludes, for example, interactions that contain inverse powers of the spatial Laplacian. This is a bit too restrictive since we know of local theories where interactions that are not manifestly local arise when integrating our non-dynamical fields, for example, the lapse and the shift of the metric in the ADM formulation of general relativity, so it is important to improve on this in the future. The resulting constraint is the manifestly local test (MLT) \cite{Jazayeri:2021fvk}:
\be \label{eq:mlt-1}
\left( \partial_{k_a} \psi_n \right) \Big|_{k_a=0} = 0\,, \quad \forall \,a=1,\dots, n\,,
\ee
where in this formula we are to think of $\psi_n$ as a function of $k_b$, $|\vec{k}_b+\vec{k}_c|$, and $\vec{k}_b \cdot \vec{k}_c$, with $b\neq c$, and these are treated as independent variables. This version of the MLT applies to any field with the same mode functions as a massless scalar or graviton. It is intuitively clear why locality gives us a constraint on $\psi_n$ at the origin of Fourier space, i.e., at $k_{a}=0 $. In position space, locality requires that correlators factorise into the product of lower-order correlators as we take the separation of clusters of points to infinity. This is the idea of cluster decomposition (see, e.g., Ref.~\cite{Weinberg:1995}). Furthermore, the connected part of the correlator is required to vanish sufficiently rapidly in this limit. Standard results from Fourier analysis then tell us that the decay of a function as the position space coordinate is taken to infinity constrains its Fourier transform at the origin. We will expand on this observation elsewhere. 

Two ways of deriving the MLT were given in Ref.~\cite{Jazayeri:2021fvk}. One way to understand Eq.~\eqref{eq:mlt-1} is as a consequence of the allowed singularity structure of wavefunction coefficients, as codified by the cosmological optical theorem \cite{COT}, together with the explicit form of the low-energy expansion of the massless bulk-to-bulk propagator. Alternatively, from a purely bulk perspective this condition follows from the following property of derivatives of the massless bulk-to-boundary propagator:
\be \label{eq:mode-derivatives}
K_{\phi}(\eta, k)=(1-i k \eta) e^{ik \eta} \implies \frac{\partial}{\partial k} \left( \frac{d^n}{d \eta^n} K_{\phi}(\eta, k) \right) \Bigg|_{k=0} =0 ,
\ee
where $\eta$ is the conformal time in the bulk of the de Sitter space.
The bulk computation of wavefunction coefficients amounts to computing nested time integrals, where the integrands are the product of derivatives of a bulk-to-boundary propagator for each external field together with contractions of spatial momenta and derivatives of bulk-to-bulk propagators (for exchange diagrams). If we keep the internal energies and contractions of spatial momenta fixed, then the MLT for wavefunction coefficients follows directly from Eq.~\eqref{eq:mode-derivatives}. For a more detailed discussion of these arguments, we refer to Ref.~\cite{Jazayeri:2021fvk}.

 
\section{The contact reconstruction formula}
\label{sec:WFfromAmp}

In this section we present a \textit{contact reconstruction formula}\footnote{Our contact reconstruction formula is not to be confused with the reconstruction formulas used to derive spinning correlators from their transverse-traceless parts, e.g., as in Ref.~\cite{Bzowski:2013sza}.} for turning contact, manifestly local $n$-point amplitudes for the graviton and any number of massless scalar fields into wavefunction coefficients at the boundary of de Sitter spacetime. 
Since amplitudes are much easier to compute, this lets us write down explicitly the EFT expansion of the wavefunction coefficients. At a technical level, our contact reconstruction formula provides an explicit solution to the MLT derived in Ref.~\cite{Jazayeri:2021fvk}. 


\subsection{Boost-breaking kinematics}
We begin by reviewing the kinematics of boost-breaking massless scalar amplitudes in four dimensions. The $n$-point kinematic variables are the $n$ energies, $k_a$, and the $n$ spatial 3-momenta, $\vec{k}_a$, which we take to be incoming. These momenta and energies satisfy the following on-shell conditions:
\begin{align}
k_1 + k_2 + \dots + k_n & =0, \\
\vec{k}_1 + \vec{k_2} + \dots + \vec{k}_n & =0, \\
\vec{k}_a \cdot \vec{k}_a - k_a^2 & =0, \quad a=1, \dots n,
\end{align}
corresponding to energy conservation, momentum conservation, and the free equation of motion. Note that we have assumed a relativistic dispersion relation, even though the interactions can break boost invariance. For wavefunction coefficients the on-shell conditions are the same except that we drop energy conservation. In that case $  k_a $ is really just the norm of the vector $  \vec{k}_a $, rather than the energy, since the energy is not a conserved quantum number in a time-dependent background. However, we still refer to $  k_a $ as the ``energy'' even on curved spacetime to facilitate the use of flat-spacetime intuition. 

A parity-even boost-breaking amplitude is an $SO(3)$-invariant function of these variables.\footnote{The constraints on boost-breaking amplitudes of spinning particles due to consistent factorisation at four points were studied recently in Refs.~\cite{PSS, Stefanyszyn:2020kay}.} It is useful to define Mandelstam-like variables as certain contractions of the spatial momenta,
\be \label{eq:Mandelstam}
s_{ab} \equiv (\vec{k}_a + \vec{k}_b)^2,
\ee
for $a,b=1, \dots, n$. These contractions are symmetric under interchanging $a$ and $b$, so before enforcing the on-shell conditions there are $n(n+1)/2$ such variables. 
For parity-odd interactions the amplitude can also depend on contractions of the momenta with the completely antisymmetric tensor, but we will restrict to parity-even interactions in this paper.

Using the free equation of motion we can always eliminate the invariants $s_{aa}$ in terms of the energies, and using energy conservation we can eliminate one of the energies in terms of the others.  Momentum conservation gives $n$ additional constraints,
\be \label{eq:sij-constraints}
\sum_{ \substack{b=1 \\ b \neq a}}^n s_{ab} = (n-4)k_a^2+\sum_{b=1}^n k_b^2, \quad a=1, \dots, n,
\ee
which can be used to eliminate a further $n$ Mandelstam variables. The amplitude then depends on $n-1$ energies and $n(n-3)/2$ Mandelstam variables. The number of independent kinematic variables is $3n-7$, so for $n \geq 5$ there are additional relations between the energies and Mandelstam variables coming from Gram identities, which follow from the fact that any four $3$-momenta cannot be linearly independent. The simplest Gram identity is the following relation, which gives a nontrivial constraint on the energies and Mandelstam variables for $n\geq 5$:
\be
\delta^{[i_1}_{j_1} \cdots \delta^{i_4]}_{j_4} k_{1,i_1} k_1^{j_1} \cdots k_{4, i_4} k_4^{j_4} =0,
\ee
where $k_{a, i}$ denotes the $i$\textsuperscript{th} component of the vector $\vec{k}_a$.
If some of the scalars are identical then the amplitude is also required to be invariant under permutations of the energies and momenta of the identical particles, up to terms that vanish on shell. In general, it is difficult to construct a basis of kinematic invariants due to the nonlinearity of the Gram identities. At four points there are no non-trivial Gram identities. In Section \ref{sec:trispectrum} we describe an explicit basis for boost-breaking four-point contact amplitudes of identical scalars. 

Now consider the particular case of a boost-breaking \textit{contact} amplitude $A_n$ for $n$ massless scalars. This amplitude is a polynomial function of the energies and  Mandelstam variables,
\be \label{eq:amp-expand}
A_n = \sum_{q=0}^{q_{\rm max}} A_n^{(q)}(k_a, s_{ab}),
\ee
where $A_n^{(q)}$ are homogeneous polynomials of order $q$ under rescaling energy and momentum, i.e., 
\be
A_n^{(q)}( \lambda k_a,  \lambda^2 s_{ab}) = \lambda^q A_n^{(q)}(k_a, s_{ab}),
\ee
and $q_{\rm max}$ is the maximum homogeneity degree, corresponding to the dimension of the most irrelevant operator that is considered at a given order in an EFT expansion. These amplitudes must be invariant under the action of the subgroup of the symmetric group $S_n$ that acts on the kinematic variables by interchanging the momenta and energies of identical particles. Since this invariance is only required to hold up to terms that vanish on shell, the permutation invariance may not be manifest in a given presentation of the amplitude.

\subsection{Wavefunctions from amplitudes}
\label{ssec:WFCfromAmps}
Given any contact amplitude $A_n$, it should be possible to find a wavefunction coefficient $\psi_n$ that reduces to this amplitude on the leading total-energy pole \cite{Maldacena:2011nz,Raju:2012zr}. The explicit formula derived in Ref.~\cite{COT} dictates that 
\be \label{eq:leading-pole}
\psi_n = (p-1)!  (iH)^{p-n-1}\frac{e_n A_n^{(p-n+3)}}{k_T^{p}} + \dots ,
\ee
where the ellipsis denotes terms with subleading total-energy poles and $A_n^{(p-n+3)}$ is the part of the amplitude that is of highest order in the energies and momenta, i.e., of order $q_{\rm max} = p-n+3$ in Eq.~\eqref{eq:amp-expand}. We will ignore the imaginary parts of the wavefunction coefficient, which are only present in the IR divergent pieces \cite{COT}.
Intuitively, the reason such a wavefunction coefficient should exist is because we can use the same interaction vertex underlying the amplitude to construct a wavefunction coefficient through a bulk computation.
Such a wavefunction coefficient will not be unique in general, since we can add wavefunction coefficients coming from contact interactions with fewer derivatives without affecting the residue of the leading $  k_{T} $ pole. In other words, the residue of the leading total-energy pole only gives the high-energy limit of the amplitude. For this reason we can also restrict to amplitudes that are homogeneous of degree $p-n+3$. We will give a closed-form algebraic expression for $\psi_n $ in terms of the amplitude for the case of contact amplitudes of massless scalars and gravitons, making some particular choice for the subleading terms.

Our tool for constructing the wavefunction from the amplitude is the MLT \cite{Jazayeri:2021fvk} given in Eq.~\eqref{eq:mlt-1}, which we repeat here for convenience,
\be \label{eq:mlt}
\left( \partial_{k_a} \psi_n \right) \Big|_{k_a=0} = 0\,, \quad \forall \,a=1,\dots, n\,.
\ee
This condition relates terms with different total-energy poles, so fixing the leading pole as in Eq.~\eqref{eq:leading-pole} will generally mandate the presence of certain subleading poles---the exception is when the amplitude is proportional to $e_n$, in which case no subleading poles are needed. The residues of these subleading poles can be fixed by writing a general ansatz and using Eq.~\eqref{eq:leading-pole} to fix the free coefficients in this ansatz, but here we give a general formula that expresses these residues directly in terms of the amplitude.
Explicitly, a wavefunction coefficient satisfying Eq.~\eqref{eq:mlt} with a prescribed leading total-energy pole as in Eq.~\eqref{eq:leading-pole} is given by the following \textit{contact reconstruction formula}:
\begin{eBox}
\be \label{eq:ampToWFnpts}
\psi_n  = (p-1)! (iH)^{p-n-1}\sum_{m=0}^{n}  \sum_{\pi \in S_n} \frac{ A_n^{(p-n+3)}\big|_{ \{ k_{\pi(j)} =0\}_{j=n-m+1}^n } \prod_{i=1}^{n-m} k_{\pi(i)} }{m! (n-m)! k_T^{p-m} \prod_{l=1}^m (p-l) },
\ee
\end{eBox}
where the second sum runs over the $n!$ permutations $\pi$ of $\{1, 2, \dots, n\}$.
A priori this formula is only valid when $p \geq n+1$, corresponding to interactions with four or more derivatives. When $p<n+1$ we can still use this formula to generate the subleading $k_T$ poles, by restricting the second sum over $  m $ to $ m\leq p-1$, but it may also be necessary to add terms analytic in the energy and/or logarithmic terms to satisfy the MLT. In the case of identical massless scalars, it can be checked that the only case for which extra terms must be added to Eq.~\eqref{eq:ampToWFnpts} is the constant amplitude (the expression for the $n$-point  wavefunction for this case is given in Eq.~\eqref{eq:nPtLog}).   

An important comment is that the amplitude $A_n^{(p-n+3)}$ used in Eq.~\eqref{eq:ampToWFnpts} must be written in a form that is manifestly symmetric under permuting the energies $k_a$ of the identical particles. This ensures that the wavefunction has the correct permutation symmetries and can be achieved by averaging over the permutations of the identical external particles (without enforcing energy and momentum conservation). The ambiguity of the subleading poles of the wavefunction coefficients comes from the fact that there are many ways to extend the amplitude away from the energy conservation constraint surface, since we can add terms proportional to the total energy without affecting the on-shell amplitude. Once we have evaluated the right-hand side of Eq.~\eqref{eq:ampToWFnpts}, we can enforce momentum conservation and this does not spoil the fact that the wavefunction satisfies the MLT because the constraints in Eq.~\eqref{eq:sij-constraints} are quadratic in the energies. 

We can prove that this expression for the wavefunction coefficient satisfies the MLT when $p \geq n+1$ by explicitly substituting it into Eq.~\eqref{eq:mlt}. It is easiest to see how this works by looking at an example. Taking $n=3$ and expanding out the sums in Eq.~\eqref{eq:ampToWFnpts}, the contact reconstruction formula gives
\begin{align} \label{eq:reconstruct-n=3}
\psi_3  & = (p-1)! i^p H^{p-4} \bigg[ \frac{ A_3^{(p)} k_1 k_2 k_3 }{ k_T^{p} } + \frac{   A_3^{(p)}\big|_{k_1 =0} k_2 k_3+ A_3^{(p)}\big|_{k_2 =0} k_1 k_3  + A_3^{(p)}\big|_{k_3 =0} k_1 k_2  }{ k_T^{p-1} (p-1) } \nonumber \\
& + \frac{  A_3^{(p)}\big|_{ k_2=k_3=0} k_1 + A_3^{(p)}\big|_{ k_1=k_3=0} k_2+ A_3^{(p)}\big|_{ k_1=k_2=0} k_3}{k_T^{p-2}  (p-1)(p-2) } + \frac{A_3^{(p)}\big|_{ k_1=k_2=k_3=0 } }{k_T^{p-3}  (p-1)(p-2)(p-3) }\bigg].
\end{align}
Now it is straightforward to check that we get zero if we take the derivative of the numerator of the first term in square brackets with respect to $k_3$, add to it the derivative of the denominator of the second term in square brackets with respect to $k_3$, and then set $k_3=0$. Similar cancellations occur for any pair of consecutive terms and for any $k_i$. Additionally, the derivative of the denominator of the first term with respect to $k_i$ and the derivative of the numerator of the last term with respect to $k_i$ both vanish after setting $k_i$=0 for any $i$, so overall the expression satisfies the MLT.\footnote{The last term in Eq.~\eqref{eq:reconstruct-n=3} vanishes, but this is a special property of scalar bispectra and so we left it in to illustrate the general argument.}
 Similar reasoning works for any $n\geq3$.
For particles with different speeds of sound we should use the corresponding amplitude and replace $k_a \rightarrow c_s^{(a)} k_a$ on the right-hand side of Eq.~\eqref{eq:ampToWFnpts} (not touching the amplitude).


\subsection{Bispectra}

In this subsection, as a simple illustration of Eq.~\eqref{eq:ampToWFnpts}, we derive cubic wavefunction coefficients and discuss how they are simply related to the primordial three-point function $  B_{3} $ of curvature perturbations $  \zeta $, a.k.a. the bispectrum, which is constrained by cosmological observations.

Assuming single-field slow-roll inflation, we can derive the bispectrum of the gauge invariant curvature perturbations $\zeta$ by a simple linear transformation of the cubic wavefunction coefficients $  \psi_{3} $ of a single scalar. By interpreting the quantum field appearing in the wavefunction as the fluctuation $  \delta \phi $ of the inflaton, the correlators of curvature perturbations are derived with the rescaling $\zeta= \delta \phi/(M_p\sqrt{2 \epsilon})$, where $  \e\equiv -\dot H/H^{2} $ is the Hubble slow-roll parameter. This relation in general contains also higher-order terms, but they are generically slow-roll suppressed and can be neglected to leading order when the inflaton self-interactions are large. The resulting $  \zeta $ bispectrum is given by
\begin{align}
B_{3}(k_{1},k_{2},k_{3})
=-\frac{H^{6}}{32\e^{3} c_{s}^{3}}\frac{\Re \psi_{3}(k_{1},k_{2},k_{3})}{(k_{1}k_{2}k_{3})^{3}}\,,
\end{align} 
where to be fully explicit we have re-inserted the inflaton speed of sound $  c_{s} $.
The constraints that the MLT puts on the bispectrum were already considered in Ref.~\cite{Jazayeri:2021fvk} by writing a general ansatz consistent with the bootstrap rules and imposing Eq.~\eqref{eq:mlt}.
Here we reproduce these results using Eq.~\eqref{eq:ampToWFnpts}. 

A generic cubic boost-breaking tree amplitude for identical massless scalars with manifestly local interactions is a sum of terms of the form
\be 
A^{(a,b)}_3 = (-i)^{2a+3b} c_{a,b} \mathbb{e}_2^a \mathbb{e}_3^b,
\ee
where $\mathbb{e}_2$ and $\mathbb{e}_3$ are the elementary symmetric polynomials in three variables and $c_{a,b}$ are real constants. 
If $b>0$, then the MLT is satisfied trivially by the leading $k_T$-pole term and the contact reconstruction formula \eqref{eq:ampToWFnpts} just yields the wavefunction coefficient as
\be
\psi^{(a,b)}_3 = (2a+3b-1)! H^{2a+3b-4}  \frac{c_{a,b}\mathbb{e}_2^a \mathbb{e}_3^{b+1}}{k_T^{2a+3b}}, \quad b>0.
\ee
For the simplest case $a=0$ and $b=1$, the amplitude $A_3^{(0,1)}= i c_{0,1}  \mathbb{e}_3$ is generated by cubic interactions with three derivatives. In de Sitter space, this leads to $\psi^{(0,1)}_3=({2c_{0,1}}/{H}) \mathcal{S}^{(3)}_3$, with the  shape function
\be \label{eq:psi33}
\mathcal{S}^{(3)}_3 \equiv \frac{ \mathbb{e}_3^{2}}{k_T^{3}} ,
\ee
which is the bispectrum shape arising from the $\dot\phi^3$ interaction.

If $b=0$ and $a>0$, subleading total-energy poles are needed to satisfy the MLT, and accordingly the wavefunction coefficient derived from the contact reconstruction formula \eqref{eq:ampToWFnpts} becomes
\be
\psi_3^{(a,0)} = (2a-1)! H^{2a-4} c_{a,0} \[ \frac{\mathbb{e}_2^a \mathbb{e}_3}{k_T^{2a}}+ \frac{ (k_1 k_2)^{a+1} +(k_1 k_3)^{a+1} +(k_2 k_3)^{a+1} }{ (2a-1)k_T^{2a-1}} \] , \quad a>0.
\ee
For a given positive integer $a$, we can write the subleading $k_T$ pole in terms of symmetric polynomials.
For example, for $a=1$ the flat-space amplitude is $A_3^{(1,0)}= - c_{1,0}\mathbb{e}_2$ and the reconstructed wavefunction coefficient is $\psi_3^{(1,0)} = ({c_{1,0}}/{H^2})\mathcal{S}^{(2)}_3 $, with the shape
\be \label{eq:psi32}
\mathcal{S}^{(2)}_3 \equiv \frac{\mathbb{e}_2 \mathbb{e}_3}{k_T^{2}}+ \frac{ \mathbb{e}_2^2 - 2 k_T \mathbb{e}_3 }{k_T} .
\ee
For $a=2$, we have a four-derivative amplitude $A_3^{(2,0)}=  c_{2,0}\mathbb{e}_2^2$ and $\psi_3^{(2,0)} =  6 c_{2,0} \mathcal{S}^{(4)}_3$, with
\be \label{eq:psi34}
\mathcal{S}^{(4)}_3  \equiv \frac{\mathbb{e}_2^2 \mathbb{e}_3}{k_T^{4}}+ \frac{ \mathbb{e}_2^3 + 3 \mathbb{e}_3^2 }{3k_T^3} - \frac{ \mathbb{e}_2 \mathbb{e}_3 }{k_T^2} .
\ee
One can easily check that these reconstructed wavefunction coefficients agree with the results in Ref.~\cite{Jazayeri:2021fvk}.

The case $a=b=0$ corresponds to the $\phi^3$ interaction with a constant flat-space amplitude $A_3^{(0,0)}=  c_{0,0}$. In de Sitter space, the wavefunction coefficient requires a log term to satisfy the MLT. This case requires adding analytic and log terms to the contact reconstruction formula, which can be fixed by explicitly applying the MLT, as in Ref.~\cite{Jazayeri:2021fvk}. The resulting shape function is given by
\be \label{eq:bispectraLog}
\mathcal{S}^{\rm log}_3 \equiv \(k_T^3 - 3 k_T \mathbb{e}_2 +3 \mathbb{e}_3\)\log(k_T/\mu) - k_T \mathbb{e}_2 +4 \mathbb{e}_3\,.
\ee
Lastly, there can also be a contribution to the bispectra arising from the local field redefinition $\phi\rightarrow\phi+\phi^2$, which leads to
\be
\mathcal{S}^{\rm local}_3 \equiv k_T^3 - 3 k_T \mathbb{e}_2 +3 \mathbb{e}_3 .
\ee

The above shape functions can be seen as building blocks for three-point correlators of identical massless scalars in manifestly local theories.  By taking their linear combinations, we can reproduce all possible bispectra in single field inflation.
For example, the well-known non-Gaussian shapes arising from the EFT of single-clock inflation \cite{EFTofI} can be written as
\be
&&B^{\dot\phi^3}_3 \propto \frac{1}{\mathbb{e}_3^3}\mathcal{S}^{(3)}_3\,, \\
&&B^{\dot\phi (\nabla\phi)^2}_3 \propto \frac{1}{\mathbb{e}_3^3}\( 12 \mathcal{S}^{(3)}_3  - 4 \mathcal{S}^{(2)}_3 + \mathcal{S}^{\rm local}_3 \) .
\ee

The contact reconstruction formula can also be applied to theories with multiple interacting scalars. As a simple example, consider the contact interaction $\alpha \dot\phi^2\sigma/2$ between two massless scalars $\phi$ and $\sigma$, which is relevant in multi-field inflation models.
The tree-level amplitude is given by $A_3= - \alpha k_1 k_2 $. By applying the contact reconstruction formula, we get
\be
\psi_{3}= \frac{\alpha}{H^2} \left( \frac{k_1k_2 \mathbb{e}_3}{k_T^2} + \frac{k_1^2k_2^2}{k_T}  \right).
\ee
For the interactions with one derivative per field, such as  $\dot\phi^2\dot\sigma$ and $\dot \phi \partial_i \phi \partial_i\sigma$, the amplitude is the same as for identical scalars,  $A_3 \sim \mathbb{e}_3 $, and the reconstructed shape function coincides with Eq.~\eqref{eq:psi33}.  
As we can see,  the wavefunction coefficients of non-identical scalars inherit the permutation symmetries of the amplitudes, which in general are not fully permutation invariant.

From these results, we can derive the three-point correlation functions of these two fields during inflation using Eq.~\eqref{eq:WFCTo3Pt}.
For instance, the correlator generated by $\alpha\dot\phi^2\sigma/2$ is given by
\be
 \langle \phi(\vec{k}_{1})\phi(\vec{k}_{2})\sigma(\vec{k}_{3}) \rangle' =-\frac{H^{6}}{4}\frac{\Re \psi_{3}(k_{1},k_{2},k_{3})}{(k_{1}k_{2}k_{3})^{3}}
=-\alpha {H^4}\frac{k_1k_2 \mathbb{e}_3+k_{T} k_1^2k_2^2}{{4}(k_{1}k_{2}k_{3})^{3}k_T^2} , 
\ee
where the overall delta function has been stripped off. For phenomenology, these results can be relevant in multi-field models, where one takes $\phi$  as the inflaton and $\sigma$ as an additional light field.
Then the above correlators with $\sigma$ may correspond to the adiabatic-isocurvature mixed non-Gaussianity if the isocurvature perturbation survives until the time of observation. 
Otherwise, these correlators can imprint the non-Gaussian signals in curvature perturbations if there are conversion processes from the light field $\sigma$ to the adiabatic modes during or after inflation.

 
\subsection{Comments on the contact reconstruction formula}
\label{ssec:comments}

In this subsection we discuss two related aspects of the contact reconstruction formula. First, we show that in general it gives a \textit{different} result from what one would have obtained by performing the direct bulk time integrations, although there are qualitative similarities. Second, we show that if we feed a Lorentz-invariant amplitude into the contact reconstruction formula, in general we do \textit{not} obtain a conformally invariant wavefunction coefficient. 

Consider a simple bulk time integral for a contact interaction that includes at most a single time derivative per field (as follows, e.g., from the Feynman rules as reviewed in Ref.~\cite{COT}),
\begin{align} \label{bulkint}
\psi_{n}&\sim -i \int d\eta \,\eta^{p-1-n} F(\vec k) \left[  \prod_{a=1}^{j} K'(k_{a},\eta)\right]\left[  \prod_{b=j+1}^{n} K(k_{b},\eta)\right] + {\rm perm.}\\
&\sim -i \int d\eta \,\eta^{p-1-n} F(\vec k) \left[  \prod_{a=1}^{j} k_{a}^{2}\eta \right]\left[  \prod_{b=j+1}^{n} (1-i k_{b}\eta)\right] e^{ik_{T}\eta}  + {\rm perm.}\,,
\end{align}
where $  F(\vec k) $ collects all the contractions of the spatial momenta coming from spatial derivatives; $  K $ is the bulk-to-boundary propagator, which is just the unnormalised de Sitter mode function; the power of $  \eta $ is such that the largest $  k_{T} $ pole has degree $  p $, as in \eqref{eq:ampToWFnpts}; and $  j $ counts the number of time derivatives. 
In Section \ref{sec:5.2}, we show explicitly that for contact interactions with at most one time derivative per field, Eq.~\eqref{eq:ampToWFnpts} gives the same result as the bulk integral \eqref{bulkint}.

However, in general the contact reconstruction formula and the explicit bulk time integral give different wavefunction coefficients when starting from the same off-shell Feynman vertex.
As a simple example, consider the interaction
\begin{align}
\L\supset \left( a^{-2}\phi'' \right)^{n}\,.
\end{align}
The corresponding amplitude is simply $  e_{n}^{2} $, namely the product of all energies squared. When we feed this into the contact reconstruction formula we get only a single term with a $  k_{T}^{-(3n-3)} $ pole because the amplitude is soft in all momenta, i.e., it vanishes when any $  k_a $ is set to zero. Conversely, the bulk time integral is more complicated and has non-vanishing $  k_{T} $ poles from $  1/k_{T}^{3n-3} $ down to $  1/k_{T}^{2n-3} $. What is happening is that the leading $  k_{T} $ pole already satisfies the MLT and so the contact reconstruction formula reconstructs $  \psi_{n} $ in the simplest possible way. We conclude that the contact reconstruction formula is in general distinct from the bulk time integral.

A second point we want to mention concerns Lorentz and conformal invariance. From a bulk point of view, if one starts from a Lorentz-invariant interaction in Minkowski space and extends it using minimal coupling to de Sitter space, then this interaction leads to a de Sitter invariant wavefunction coefficient. Once this wavefunction is pushed to the future (conformal) boundary, the de Sitter isometries can be interpreted as the Euclidean conformal group in one lower dimension. Indeed, conformal invariance has been extensively used to derive correlators and wavefunction coefficients \cite{Maldacena:2011nz,Antoniadis:2011ib,Creminelli:2011mw,Mata:2012bx,Kehagias:2012pd,Kehagias:2012td,Kundu:2014gxa,Kundu:2015xta,Pajer:2016ieg,CosmoBootstrap1,CosmoBootstrap2,CosmoBootstrap3}. It is then natural to ask whether the contact reconstruction formula gives a conformally invariant $  \psi_{n} $ if we start from a Lorentz-invariant amplitude. As we now show, this is not the case in general. 

As a simple example, consider the three-point function induced by the cubic interaction
\begin{align}
\L \supset  \phi  \partial_{\mu}\partial_{\nu}\phi \partial^{\mu}\partial^{\nu}\phi \,,
\end{align} 
which can be field redefined to a boundary term.
The off-shell amplitude, i.e., the Feynman vertex obtained by varying the Lagrangian, can be written as
\begin{align}
A_{3}\propto k_{T}^{2}e_{2}-\frac{3}{8}k_{T}^{4}\,,
\end{align}
which indeed vanishes at $  k_{T}=0 $. Even though $A_{3}$ vanishes on shell, the contact reconstruction formula would have to transform it into a conformally invariant wavefunction coefficient if it does so to Lorentz-invariant amplitudes. However, feeding this $  A_{3} $ into the contact reconstruction formula we find that the leading $  k_{T} $ pole vanishes, as it should, but the subleading $  k_{T} $ poles do not. One can check explicitly that the resulting expression is not annihilated by the generators of special conformal transformations.\footnote{The same holds for the analogous quartic interaction $ (\phi \partial_{\mu}\partial_{\nu}\phi)^2  $, which has a non-vanishing amplitude.} This can also be seen by the following argument. There are only two conformally invariant cubic wavefunction coefficients---equivalently, cubic scalar correlators in a Euclidean conformal field theory (CFT) for $  \Delta =3 $: one has a logarithmic term $  \log k_{T} $, corresponding to the well-known $  \left( x_{12}x_{23}x_{13} \right)^{-3}$ term in position space (see, e.g., \cite{Creminelli:2011mw}), and the other is the local non-gaussianity, which is finite as $  k_{T}\to 0 $ and corresponds to a contact term in position space. Neither of these shapes have a $  k_{T} $ pole and so the result of the contact reconstruction formula is not conformally invariant. 
To summarise, if we start from a Lorentz-invariant amplitude, the result of the contact reconstruction formula is not in general conformally invariant. 

 
\section{All contact trispectra in the EFT of inflation}
\label{sec:trispectrum}

In this section we apply the contact reconstruction formula to the four-point function. 
We first write down a general boost-breaking four-particle contact amplitude and then derive the phenomenologically relevant trispectra from scalar contact interactions during inflation. An infinite number of  de Sitter invariant four-point contact correlators for conformally coupled and massless scalars were constructed using bootstrap methods in Ref.~\cite{CosmoBootstrap1}, namely those that arise from integrating out the single exchange of a heavy scalar.


\subsection{Four-particle amplitudes}
\label{ssec:4ptHilbert}

We want to find an expression for the general four-particle boost-breaking amplitude of identical massless scalars. First we find the Hilbert series for the underlying polynomial ring. See Refs.~\cite{Henning:2015daa, Henning:2017fpj} for introductions to the use of Hilbert series techniques to enumerate amplitudes. Following the general discussion of kinematics in Section~\ref{sec:WFfromAmp}, we can write a general parity-even four-particle contact amplitude of identical massless scalars as an element of the following polynomial ring:
\be
R \equiv \left[\frac{\mathbb{C}[k_1,k_2,k_3,k_4, \vec{k}_1, \vec{k}_2, \vec{k}_3, \vec{k}_4]}{\langle \{\vec{k}_a \cdot \vec{k}_a-k_a^2\}_{a=1}^4, \, \sum_{a=1}^4 k_a,  \, \sum_{a=1}^4 \vec{k}_a \rangle} \right]^{{\rm SO}(3) \times S_4},
\ee
where the ideal in the denominator enforces the on-shell conditions and the superscript ${\rm SO}(3) \times S_4$ means that we restrict to combinations of the kinematic variables that are singlets under rotations and permutations.
Now passing to the Mandelstam variables defined in Eq.~\eqref{eq:Mandelstam} and using some of the constraints to eliminate $s_{11}, \dots,  s_{44}$, $s_{23}$, $s_{24}$ and $s_{34}$, we get 
\be
R =\left[\frac{\mathbb{C}[k_1,k_2,k_3,k_4, s_{12}, s_{13}, s_{14}]}{\langle  \sum_{a=1}^4 k_a, s_{12}+s_{13}+s_{14}-\sum_{a=1}^4 k_a^2 \rangle} \right]^{S_4} .
\ee
We define $R'$ as the ring without the quotient,
\be
R' \equiv \mathbb{C}[k_1,k_2,k_3,k_4, s_{12}, s_{13}, s_{14}]^{S_4} .
\ee
The Hilbert series of $R'$ can be found with Molien's formula,
\begin{align}\label{moliens}
\mathcal{H}_{R'}(t; r) & = \frac{1}{4!} \sum_{\pi \in S_4} \frac{1}{\det (1-  F M_{\pi} )},
\end{align}
where $M_{\pi}$ are $7 \times 7$ matrices that encode the linear action of the permutations $\pi\in S_4$  on the variables $k_1,k_2,k_3,k_4, s_{12}, s_{13}, s_{14}$. For example, the permutation represented by the 2-cycle $\pi=(3,4)$ permutes the momenta $\vec k_{3} $ and $  \vec k_{4} $ and therefore acts as the simultaneous exchange $  k_{3} \leftrightarrow k_{4} $ and $  s_{13} \leftrightarrow s_{14} $. Similarly, $\pi=(1,2)$ corresponds to the permutation of $  \vec k_{1} $ and $  \vec k_{2} $ and acts as $  k_{1} \leftrightarrow k_{2} $ and $  s_{13} \leftrightarrow s_{14} $, and so on.
The matrix $  F $ in the denominator of Eq.~\eqref{moliens} is
\be
F\equiv {\rm diag} \! \left(  t, t,  t, t, r t^2, r t^2, r t^2 \right),
\ee
where the parameter $t$ keeps track of the total power of energy and momentum, while $r$ tracks the powers of the variables $s_{ab}$. From Eq.~\eqref{moliens} we get
\be
\mathcal{H}_{R'}(t; r) = \frac{1+r t^4+r(r+1) t^6+r^2 t^8+r^3 t^{12}}{(1-t)(1-t^2)(1-t^3)(1-t^4)(1-r t^2) (1-r^2t^4)(1- r^3t^6)}.
\ee
The remaining constraints are enforced by removing the primary generators of order one and two, corresponding to $k_T$ and $s_{12}+s_{13}+s_{14}$.  To get the Hilbert series for $R$ we thus multiply $\mathcal{H}_{R'}(t; r)$ by $(1-t)(1-rt^2)$, so we get
\be \label{eq:HilbertFinal}
\mathcal{H}_{R}(t; r) = \frac{1+r t^4+r(r+1) t^6+r^2 t^8+r^3 t^{12}}{(1-t^2)(1-t^3)(1-t^4)(1-r^2t^4)(1- r^3t^6)}.
\ee
Setting $r=1$, the first few terms in the series expansion around $t=0$ are
\be \label{eq:4PtHilbert}
\mathcal{H}_{R}(t) \equiv \mathcal{H}_{R}(t; 1) = 1+t^2+t^3+4t^4+t^5+8t^6+4t^7+14t^8+8t^9+24t^{10}+\dots,
\ee
so the number of contact amplitudes with $q$ powers of energy and momentum is $1,0,1,1,4, \dots$ for $q=0,1,2, \dots$. Below we will give the first few corresponding interactions.

The Hilbert series \eqref{eq:HilbertFinal} indicates that the ring $R$ admits a Hironaka decomposition with five primary generators and five non-trivial secondary generators. The number of primary generators can be read off from the number of factors in the denominator counted with multiplicities, i.e., factors to the power of $  n $ count as $  n $ generators (in our case the exponent is one for all factors). The total number of secondary generators comes from setting $t=r=1$ in the numerator.
The Hilbert series also tells us the scaling of each generator with $  k_{a} $ and $  s_{ab} $. For example, the term $ (1-r^2t^4)$ in the denominator implies that there is one primary generator scaling as $  s^{2} $.
The numerator fixes the same scaling for the secondary generators. For example, the monomial $ r^2 t^8 $ implies that there is one secondary generator scaling as $  s^{2} k^{4} $, while the monomial $ r t^4 $ indicates the scaling $  s k^{2} $.

We can find an explicit set of primary and secondary generators using the program \texttt{Macauley2} with the package \texttt{InvariantRing} \cite{invariantring}. The five primary generators are $e_2$, $e_3$, $e_4$, and
\begin{align} \label{primary}
E_2 & \equiv s_{12} s_{13} + s_{12} s_{14}+s_{13} s_{14}, \\
E_3 & \equiv s_{12} s_{13} s_{14},
\end{align}
and the five non-trivial secondary generators can be written as
\begin{align}
S_1 &  \equiv (k_1 k_2  + k_3 k_4 )s_{12} + (k_1 k_3  + k_2 k_4) s_{13} +( k_2 k_3 + 
 k_1 k_4 )s_{14}, \\
S_2 & \equiv (k_1 k_2  + k_3 k_4 )s_{12}^2 + (k_1 k_3 + k_2 k_4 )s_{13}^2 + (k_2 k_3  + 
 k_1 k_4 )s_{14}^2, \\
 S_3 & \equiv (k_1^3 k_2 + k_1 k_2^3 + k_3^3 k_4 + k_3 k_4^3) s_{12} + (k_1^3 k_3 + 
 k_1 k_3^3  + k_2^3 k_4 + k_2 k_4^3 )s_{13} \nonumber \\
 &+ (k_2^3 k_3  + k_2 k_3^3  +  k_1^3 k_4 + k_1 k_4^3) s_{14}, \\
  S_4 & \equiv (k_1^3 k_2  + k_1 k_2^3  + k_3^3 k_4  + k_3 k_4^3 )s_{12}^2 + 
( k_1^3 k_3  + k_1 k_3^3  + k_2^3 k_4  + k_2 k_4^3 )s_{13}^2 \nonumber \\
 &+(k_2^3 k_3  + k_2 k_3^3  + k_1^3 k_4  + k_1 k_4^3 )s_{14}^2, \\
  S_5 & \equiv  (k_1^3 k_2^2 k_3  + k_1^2 k_2^3 k_4  + 
 k_1 k_3^3 k_4^2  + k_2 k_3^2 k_4^3 )s_{12}^2 s_{13} + ( k_1^3 k_2 k_3^2  + k_1^2 k_3^3 k_4  +  k_1 k_2^3 k_4^2  + k_2^2 k_3 k_4^3) s_{12} s_{13}^2\nonumber \\
 & +  (k_1^2 k_2^3 k_3  + k_1^3 k_2^2 k_4 + 
 k_2 k_3^3 k_4^2  + k_1 k_3^2 k_4^3 )s_{12}^2 s_{14} + ( k_1^2 k_2 k_3^3 + k_1^3 k_3^2 k_4 + 
 k_2^3 k_3 k_4^2  + k_1 k_2^2 k_4^3 )s_{13}^2 s_{14}  \nonumber \\
 & +  (k_1 k_2^3 k_3^2  + k_2^2 k_3^3 k_4  +  k_1^3 k_2 k_4^2 + k_1^2 k_3 k_4^3 )s_{12} s_{14}^2 + ( k_1 k_2^2 k_3^3  + k_2^3 k_3^2 k_4  + 
 k_1^3 k_3 k_4^2 + k_1^2 k_2 k_4^3 )s_{13} s_{14}^2 . \label{secondary}
\end{align}
These scalings agree with what is dictated by the Hilbert series. Defining $S_0 \equiv1$, we can  write a general boost-breaking quartic contact amplitude of identical massless scalars as
\be\label{A4}
A_4 = \sum_{i=0}^5 P_i(e_2, e_3, e_4, E_2, E_3) S_i,
\ee
where $P_i$ are general polynomial functions of their arguments.


\subsection{Four-point wavefunction coefficients}

Now that we have a generic expression for the four-particle amplitude $ A_{4} $, we can apply the contact reconstruction formula given in Section \ref{sec:WFfromAmp} to derive quartic contact wavefunction coefficients $  \psi_{4} $ to any desired order in the derivative expansion starting from $  p \geq 2 $. Previous work on the trispectrum from a bulk perspective includes Refs.~\cite{Huang:2006eha,Seery:2008ax,Arroja:2008ga,Leblond:2010yq,Bartolo:2010di,Creminelli:2011mw,Renaux-Petel:2013ppa,Smith:2015uia}.

The contact wavefunction coefficients are simply related to the inflationary trispectra by
\begin{align} \label{B4psi4}
B_{4}=-2\frac{\Re \psi_{4}}{\prod_{a=1}^{4}2\Re \psi_{2}(k_{a})}=-\frac{H^{8}}{128\e^{4} c_{s}^{4}}\frac{\Re \psi_{4}}{(k_{1}k_{2}k_{3}k_{4})^{3}}\, .
\end{align}
The trispectra $  B_{4} $ are one of the major observational targets of primordial non-Gaussianity being tested by cosmological surveys.
Here we write the contact wavefunction coefficients $\psi_{4}$ produced by the contact reconstruction formula \eqref{eq:ampToWFnpts}  for $p\leq 6$. In Appendix \ref{App:A} we present a more laborious version of this calculation by writing a general bootstrap ansatz and then solving the MLT, in the same way as was done for the cubic wavefunction $  \psi_{3} $  in Ref.~\cite{Jazayeri:2021fvk}. 


The explicit reconstructed wavefunction coefficients take the following form:
\begin{itemize}
\item $p=0$ and $p=1$. These two lowest-order terms are not captured by the contact reconstruction formula. We simply collect the results from the explicit MLT solution presented in  Appendix \ref{App:A},
\begin{align}
 \psi^{(0)}_4 & = 3e_3 - { 3 k_Te_2}  +{k_T^3},  \\
 \psi^{(1)}_4 & = -3 \frac{e_4}{k_T} +4 e_3 - k_T e_2+ \( k_T^3 - 3 k_T e_2 +3  e_3 \) \log\( \frac{k_T}{\mu}\),
\end{align}
where  $\psi^{(0)}_4 =\psi^{\rm local}_4$ corresponds to the contribution  from the local field redefinition $\phi \rightarrow \phi + \phi^3$ and $\psi^{(1)}_4$ is the four-point correlator arising from the $\phi^4$ interaction.

\item $ p=2$. There is no four-particle amplitude from a one-derivative contact interaction, as the only boost-breaking operator $\dot\phi\phi^3$ can be written as a total derivative. Thus no wavefunction coefficient  arises at this order.

\item $p=3$. There is one possible amplitude from two-derivative interactions,
\be
A_4^{(2)} = -\lambda^{(2)} e_2 ,
\ee
where $\lambda^{(2)}$ is a coupling constant.
It is generated, for example, by the boost-breaking quartic interactions $\dot\phi^2\phi^2$, $\partial_i\phi \partial_i \phi \phi^2$, $\ddot\phi \phi^3$, and $ (\partial_i^2 \phi)\phi^3$.
The contact reconstruction formula leads to $\psi^{(3)}_4 = 2 \lambda^{(2)} H^{-2} \mathcal{S}^{(3)}_4 $, with the trispectrum shape
\begin{align} \label{psi43r}
\mathcal{S}^{(3)}_4 & \equiv \frac{e_2 e_4}{k_T^3} + \frac{\sum_{S_4}\[k_1k_2k_3 (k_1k_2+k_1k_3+k_2k_3)\]}{3!\times 2k_T^2} + \frac{\sum_{S_4}k_1^2k_2^2}{2!\times 2!\times 2k_T} \nn\\
&=  \frac{e_2 e_4}{k_T^3} + \frac{e_2 e_3}{2k_T^2} + \frac{e_2^2-e_4}{ 2k_T} - e_3 ,
\end{align}
where in the second step we write the result completely in terms of elementary symmetric polynomials. 

\item $p=4$. For interactions with  three derivatives, there is one scattering amplitude, 
\be
A_4^{(3)} = i \lambda^{(3)}e_3 ,
\ee
which arises from interactions such as $ \dot\phi^3\phi$,~ $\dot\phi\ddot\phi \phi^2 $,~$ \dddot\phi \phi^3 $,~$\dot\phi \partial_i\phi \partial_i \phi \phi  $,~$\partial_i\dot\phi \partial_i \phi \phi^2 $,~$ \dot\phi(\partial_i^2 \phi)\phi^2$ and $(\partial_i^2 \dot\phi)\phi^3$.
By applying the contact reconstruction formula, we find $\psi^{(4)}_4 = 6\lambda^{(3)}H^{-1}\mathcal{S}^{(4)}_4$, with the shape function
\be \label{psi44r}
\mathcal{S}^{(4)}_4 \equiv  \frac{e_3 e_4}{k_T^4} + \frac{e_3^2 - 2e_2e_4}{3k_T^3}  .
\ee

\item $p=5$. Four possible amplitudes arise at this order, 
\be
A_4^{(4a)} = \lambda^{(4a)} e_4~ , ~~~~~ A_4^{(4b)} = \lambda^{(4b)}e_2^2 ~, ~~~~~ A_4^{(4c)} = \lambda^{(4c)}E_2 ~, ~~~~~ A_4^{(4d)} =\lambda^{(4d)} S_1 .
\ee
The first one corresponds to the $\dot\phi^4$ operator, while the rest are generated by combinations of other four-derivative interactions, e.g., $\dot\phi\dddot\phi \phi^2$, $\ddot\phi \ddot\phi \phi^2$, $\dot\phi^2 \partial_i\phi \partial_i \phi$, $(\partial_i\phi \partial_i \phi)^2$, and $\ddot\phi \partial_i\phi \partial_i \phi \phi $.
For these amplitudes, the contact reconstruction formula yields the quartic wavefunction coefficients
\begin{align}
\psi^{(5a)}_4 &= 24 \lambda^{(4a)} \mathcal{S}^{(5a)}_4 ,& \psi^{(5b)}_4 &= 24 \lambda^{(4b)} \mathcal{S}^{(5b)}_4 ,\\
\psi^{(5c)}_4 &= 24 \lambda^{(4c)} \mathcal{S}^{(5c)}_4 , &
\psi^{(5d)}_4 &= 24 \lambda^{(4d)} \mathcal{S}^{(5d)}_4 ,
\end{align}
where the corresponding shape functions are defined by
\begin{align} \label{psi45ar}
 \mathcal{S}^{(5a)}_4 & \equiv \frac{ e_4^2}{k_T^5}  ,\\
\label{psi45br}
 \mathcal{S}^{(5b)}_4 &\equiv 
 \frac{ e_2^2 e_4}{k_T^5} 
+ \frac{ e_2^2 e_3 +3 e_3 e_4}{4k_T^4} + \frac{ 3e_3^2+ e_2^3- 18 e_2 e_4}{12k_T^3} - \frac{ e_2 e_3}{4k_T^2} + \frac{e_4}{4k_T} , \\
\label{psi45cr}
 \mathcal{S}_4^{(5c)} &\equiv 
  \frac{E_{2}e_4}{k_T^5} + \frac{E_{2}e_3}{4k_T^4}+ \frac{E_{2}e_2}{12k_T^3} + \frac{E_{2}}{12k_T} ,\\
 \label{psi45dr}
 \mathcal{S}_4^{(5d)} &\equiv
 \frac{S_1 e_4}{k_T^5} + \frac{S_1 e_3}{4k_T^4}+ \frac{2e_4e_2- S_1e_2 - S_3}{12k_T^3}+ \frac{e_2e_3}{6k_T^2}+ \frac{S_1 - e_4}{12k_T} -\frac{e_3}{12} .
\end{align}

\item $p=6$. Here we have one possible four-particle amplitude   
\be
A_4^{(5)} = -i \lambda^{(5)} e_2 e_3 ,
\ee
coming from the five-derivative interaction $\partial_i \dot{\phi} \partial^i \phi \dot{\phi}^2$. From this amplitude, we get the wavefunction coefficient $\psi^{(6)}_4 = 5!\lambda^{(5)}  H \mathcal{S}^{(6)}_4 $, where the shape given by the contact reconstruction formula is
\begin{align} \label{psi46r}
\mathcal{S}^{(6)}_4 
&\equiv
 \frac{ e_2e_3 e_4}{k_T^6} 
+ \frac{ e_2 e_3^2 +4 e_4^2 - 2 e_2^2e_4}{5k_T^5} - \frac{ e_3e_4}{5k_T^4}.
\end{align}
\end{itemize}

Taking the interacting field as the inflaton, the above results are directly related to the observables of primordial non-Gaussianity.
More precisely, all contact trispectra from the EFT of single-clock inflation  can be written as some linear combination of these reconstructed shape functions. 
Let us take the $P(X, \phi)$ theory as an example, where quartic interactions have at most four derivatives (and thus $p=5$). The leading contributions to the trispectrum in the limit of small sound speed have been calculated in Ref.~\cite{Huang:2006eha}, which correspond to the following linear combinations:
\begin{align}
& B_4^{\dot\phi^4} \sim  \frac{1}{e_4^3}  \mathcal{S}^{(5a)}_4 , \label{phidot4} \\
& B_4^{\dot\phi^2(\partial_i\phi)^2} \sim \frac{1}{e_4^3}  \( 4\mathcal{S}^{(5a)}_4 +\mathcal{S}^{(5d)}_4 -\frac{3}{2} \mathcal{S}^{(4)}_4 \) , \\
& B_4^{(\partial_i\phi)^4} \sim \frac{1}{e_4^3} \( 2 \mathcal{S}^{(5a)}_4 +  \mathcal{S}^{(5b)}_4-  \mathcal{S}^{(5c)}_4 -\frac{9}{4} \mathcal{S}^{(4)}_4+ \frac{1}{6}\mathcal{S}^{(3)}_4\) .
\end{align}

Using the contact reconstruction formula we can easily find  trispectra shapes from higher-derivative interactions without doing a bulk computation. 
For instance, if we consider quartic interactions with five derivatives during inflation,
a new contribution to the trispectrum comes from Eq.~\eqref{psi46r}.
Then by Eq.~\eqref{B4psi4} the final trispectrum can be written as 
\be
B_4=  \frac{H^{8}}{128\e^{4} c_{s}^{4}} 
\frac{120 H \lambda^{(5)} }{e_4^{3}} \[\frac{ e_2e_3 e_4}{k_T^6} 
+ \frac{ e_2 e_3^2 +4 e_4^2 - 2 e_2^2e_4}{5k_T^5} - \frac{ e_3e_4}{5k_T^4}\] + \dots,
\ee
where the dots represent other manifestly local shapes with $p<6$. By using a basis of amplitudes we obtain all possible trispectra on the boundary. 
Operators with more derivatives are expected to be suppressed compared to operators with fewer derivatives in the absence of fine tuning. 
Furthermore, inflationary models are also constrained by non-linearly realised boosts, which we have not discussed here. These non-linear symmetries constrain the relative size of different operators. For example, in single field inflation one expects that only $  \dot\phi^{4} $ in Eq.~\eqref{phidot4} can be the leading non-Gaussian signal \cite{Smith:2015uia}.


\subsection{Counting amplitudes and wavefunctions}
\label{ssec:}

We now consider the problem of counting amplitudes and wavefunction coefficients. We define $c(q)$ as the number of quartic boost-breaking contact amplitudes for identical massless scalars with exactly $q$ powers of energy and momentum, so we can write the Hilbert series \eqref{eq:4PtHilbert} as
\be
\mathcal{H}_{R}(t) = \sum_{q=0}^{\infty} c(q) t^q .
\ee
We would like to find a closed-form expression for $c(q)$. We can find a reasonably compact expression using the partial-fractions method described in Ref.~\cite{Beck2004}. The idea is to write
\be
c(q) = \frac{\mathcal{H}_{R}(t)}{t^q}\Big|_{\rm constant \, part}.
\ee
To find the constant part of $\mathcal{H}_{R}(t)/t^q$ we can expand it into partial fractions, discard the terms with poles at $t=0$, and then set $t=0$. In our case $\mathcal{H}_{R}(t)$ has poles at $\pm 1$, $\pm i$, $\pm i^{2/3}$, and $\pm i^{4/3}$, and the partial fraction decomposition takes the following form:
\begin{align}
\frac{\mathcal{H}_{R}(t)}{t^q} & =\frac{a_1}{1+t}+\frac{a_2}{(1+t)^2}+\frac{a_3}{(1+t)^3}+\frac{a_4}{(1+t)^4}+\frac{a_5}{(1+t)^5} +\frac{a_6}{1-t} +\frac{a_7}{(1-t)^2}+\frac{a_8}{(1-t)^3} \nonumber \\
&+\frac{a_9}{(1-t)^4}+\frac{a_{10}}{(1-t)^5} +\frac{a_{11}}{t+i}+\frac{a_{12}}{(t+i)^2}+\frac{a_{13}}{t-i}+\frac{a_{14}}{(t-i)^2}   +\frac{a_{15}}{t-i^{2/3}}+\frac{a_{16}}{(t-i^{2/3})^2} \nonumber \\
&+\frac{a_{17}}{t+i^{2/3}}+\frac{a_{18}}{(t+i^{2/3})^2} +\frac{a_{19}}{t-i^{4/3}}+\frac{a_{20}}{(t-i^{4/3})^2}+\frac{a_{21}}{t+i^{4/3}}+\frac{a_{22}}{(t+i^{4/3})^2} + \sum_{k=1}^q \frac{b_k}{t^k} ,
\end{align}
where $a_i$ and $b_i$ are constants depending on $q$.
The constants $a_i$ can be fixed by computing residues. For example, 
\be
a_{12} = {\rm Res} \left( \frac{(t+i)\mathcal{H}_{R}(t)}{t^q}, -i \right)=- \frac{i^{q}(1+i)}{64}.
\ee
The others can be found similarly.
The constants $b_i$ do not contribute to the constant term, so we do not need them explicitly. We can now discard the $b_i$ terms and set $t=0$ to get the following closed-form expression for $c(q)$:
\begin{align}
c(q) & = \frac{e^{-i \pi  q} (2 q+7) (2 q (q+7)+55)}{1536}+\frac{6 q (q+7) (q (q+7)+64)+4103}{13824} \nonumber \\
&+\frac{1}{36} \left(3 \cos
   \left(\frac{\pi  q}{3}\right)-\sqrt{3} \sin \left(\frac{\pi  q}{3}\right)\right)+\frac{1}{32} \left((q+5) \cos
   \left(\frac{\pi  q}{2}\right)-(q+2) \sin \left(\frac{\pi  q}{2}\right)\right) \nonumber \\
&+\frac{1}{108} \left(\sqrt{3} (2 q+5)
   \sin \left(\frac{2 \pi  q}{3}\right)+(6 q+23) \cos \left(\frac{2 \pi  q}{3}\right)\right).
\end{align}
This grows asymptotically like $c(q) \sim q^4/2304$.

 
\section{Higher-point functions}\label{sec:5}
In this section we briefly discuss how our results generalise to higher-point contact amplitudes and wavefunction coefficients. We first explain how to find the Hilbert series for higher-point contact amplitudes of identical massless scalars and then explicitly show how the contact reconstruction formula gives the expected higher-point wavefunction coefficients in certain cases.

\subsection{Higher-point Hilbert series}
The derivation of the Hilbert series for four-point amplitudes given in Section \ref{ssec:4ptHilbert} does not generalise to higher points. One way that four points is special can be seen from the factor of $n-4$ in Eq.~\eqref{eq:sij-constraints}. Another issue is that in four dimensions Gram identities become important at five points and beyond. We will not deal with this second issue, but instead we show how to derive the Hilbert Series for $n$-point amplitudes ignoring Gram identities. This means that the counting will be too large in four dimensions, although this would be the correct counting in dimensions greater than $n-1$. 

Following the discussion of kinematics in Section~\ref{sec:WFfromAmp}, we can write the polynomial ring that generates contact $n$-point boost-breaking amplitudes for identical massless scalars as
\be
R_n \equiv \left[\frac{\mathbb{C}[k_1, \dots ,k_n, s_{12}, \dots , s_{n-1 \, n}]}{\langle  \sum_{a=1}^4 k_a, \{\sum_{b \neq a} s_{ab} -(n-4)k_a^2-\sum_b k_b^2 \}_{a=1}^n \rangle} \right]^{S_n} ,
\ee
where the numerator depends on the $n$ energies $k_a$ with $a=1, \dots, n$ and the $n(n-1)/2$ variables $s_{ab}$ with $1\leq a < b \leq n$. Now for $n\geq 5$ we can enforce the constraints from momentum conservation by eliminating $k_a^2$, so we can write
\be
R_n =\left[\frac{\mathbb{C}[k_1, \dots ,k_n, s_{12}, \dots , s_{n-1 \, n},]}{\langle  \sum_{a=1}^4 k_a, k_1^2, \dots, k_n^2 \rangle} \right]^{S_n} .
\ee
Now define $R'_n$ to be the ring $R_n$ without the quotient by the ideal,
\be
R'_n \equiv \mathbb{C}\left[k_1, \dots, k_n, s_{12}, \dots,  s_{n-1 \, n}\right]^{S_n} .
\ee
The Hilbert series of $R'_n$ can be found using Molien's formula,
\begin{align}
\mathcal{H}_{R_n'}(t; r_1, \dots, r_n) & = \frac{1}{n!} \sum_{\pi \in S_n} \frac{1}{\det (1-  F_n M_{\pi} )},
\end{align}
where $M_{\pi}$ are $n(n+1)/2 \times n(n+1)/2$ matrices that encode the linear action of the permutations $\pi\in S_n$  on the variables $k_a, s_{ab}$, and
\be
F_n \equiv {\rm diag} \!\left(  r_1 t, \dots, r_n t, t^2,  \dots, t^2 \right).
\ee
The variable $t$ keeps track of the total combined power of energy and momenta, while the variables $r_a$ count the powers of $k_a$ for $a=1, \dots, n$.

To get the Hilbert series for $R_n$ we must enforce the remaining constraints. This can be achieved by extracting the part of $\mathcal{H}_{R_n'}(t; r_1, \dots, r_n)$ that is at most linear in each of the variables $r_a$ and then multiplying by $1-t$ to account for energy conservation. Altogether this gives
\begin{align}
\mathcal{H}_{R_n}(t) & =(1-t)\sum_{j=0}^n \frac{n!}{j!(n-j)!}\frac{\partial^j \mathcal{H}_{R_n'}(t; r_1, \dots, r_n)}{\partial r_1 \dots \partial r_j }\bigg|_{r_a =0},
\end{align}
where we have used the symmetry of $\mathcal{H}_{R_n'}(t; r_1, \dots, r_n)$ in the variables $r_a$.  This also gives the correct Hilbert series for $n=4$ by a similar argument with $\vec{k}_a\cdot \vec{k}_b$ replacing $s_{ab}$.

For $n=5$ we get the Hilbert series
\be
\mathcal{H}_{R_5}(t) = \frac{N(t)}{(1 - t^2) (1 - t^3) (1 - t^4)^2 (1 - t^5) (1 - t^6) (1 - t^8) (1 - 
   t^{10}) (1 - t^{12})},
\ee
where
\begin{align}
N(t)&\equiv 1 + t^4 + 2 t^5 + 3 t^6 + 5 t^7 + 7 t^8 + 8 t^9 + 11 t^{10} + 13 t^{11} + 
 20 t^{12} + 24 t^{13} + 29 t^{14}\nonumber \\ & + 34 t^{15} + 38 t^{16} 
 + 40 t^{17}  +  43 t^{18} + 49 t^{19} + 48 t^{20} + 51 t^{21} + 46 t^{22} + 45 t^{23} + 
 39 t^{24}  \nonumber \\
  & + 38 t^{25} + 31 t^{26} + 27 t^{27} + 21 t^{28} + 14 t^{29}+ 14 t^{30} + 7 t^{31} + 5 t^{32} + 3 t^{33} + 3 t^{34}.
\end{align}
This shows that, even ignoring Gram identities, it can become quite complicated to construct a basis beyond four points.
The series expansion around $t=0$ is
\be
\mathcal{H}_{R_5}(t) = 1 + t^2 + t^3 + 4 t^4 + 4 t^5 + 9 t^6 + 12 t^7 + 25 t^8 + 32 t^9 + 
 57 t^{10} +\dots,
 \ee
 which agrees with the counting we find by explicitly constructing five-point contact amplitudes with up to 10 derivatives.
 
\subsection{Higher-point wavefunction coefficients}
\label{sec:5.2}

In this subsection, we show how the contact reconstruction formula, {suitably interpreted}, agrees with the bulk integration for any $  n $-point wavefunction coefficient corresponding to contact interactions with at most one time derivative per field and an arbitrary number of spatial derivatives.

We start from the following off-shell amplitude:
\be \label{Vn}
A_n \sim  F(\vec k) \prod_{a=1}^j i k_a + {\rm perm.} \, ,
\ee
where $j$ is the number of fields with a single time derivative and $  F $ collects the contractions corresponding to all spatial derivatives.
Assuming that the above vertex comes from interactions with $p-n+3$ derivatives, where $p-n+3\geq 4$, we can write the corresponding wavefunction coefficient using Eq.~\eqref{eq:ampToWFnpts}  as
\begin{small}
\begin{align} \label{reconstr-psin}
\psi_n & \sim   \frac{ A_n e_n}{ k_T^{p}} +  \frac{1}{ k_T^{p-1}(p-1)}
 \left[ \sum_{a=j+1}^{n} F(\vec k) \frac{e_n}{k_a}\prod_{c=1}^j i k_c+ {\rm perm.}\right] +\dots\nn\\
& +  \frac{1}{ k_T^{p-m} \prod_{l=1}^m(p-l)}
\left[ F(\vec k)  \sum_{a_1=j+1}^{n} \sum_{a_2=a_1+1}^{n} \dots \sum_{a_m=a_{m-1}+1}^{n} 
\frac{e_n}{k_{a_1}k_{a_2}\dots k_{a_{m}}}\prod_{c=1}^j i k_c+ {\rm perm.}\right] + \dots ,
\end{align}
\end{small}where there are $n-j+1$ nonzero terms in this expression, from $m=0$ to $m=n-j$, {and  we treat any $\vec{k}_a \cdot \vec{k}_a$ terms in $F$ as independent of $k_a$.}

Next, let us consider the explicit bulk computation for this type of interaction, which is presented in Eq.~\eqref{bulkint}. The time integral can be written as 
\begin{align}
\psi_{n}
&\sim (-i)^{n+1} \int_{-\infty(1-i\epsilon)}^0 d\eta \,\eta^{p-1} e_n F(\vec k)  \left(  \prod_{a=1}^{j} ik_{a}  \right)  \left[  \prod_{b=j+1}^{n} \(1+ \frac{i}{ k_{b}\eta}\)\right] e^{ik_{T}\eta} {+{\rm perm.}}\, .
\end{align}
Expanding the terms in the second pair of brackets, the integral can be schematically written as $\sum_{m=0}^{n-j} C_m(k) \int d \eta \eta^{p-1-m}e^{ik_{T}\eta}$. Solving it explicitly, assuming that $p-n+3\geq 4$, we get 
\begin{small}
\begin{align}
\psi_{n}
&\sim 
e_n F(\vec k) \left(  \prod_{a=1}^{j} ik_{a}  \right) \left[  \frac{1}{k_T^p}
+ \frac{1}{k_T^{p-1}(p-1)}\sum_{a=j+1}^n\frac{1}{k_a} +\dots\right.\nn \\
& \left.+ \frac{1}{ k_T^{p-m} \prod_{l=1}^m (p-l)}
\( \sum_{a_1=j+1}^{n} \sum_{a_2=a_1+1}^{n} \dots \sum_{a_m=a_{m-1}+1}^{n}
\frac{1}{k_{a_1}k_{a_2}\dots k_{a_{m}}}\) + \dots
\right] { +{\rm perm.}}\, ,
\end{align}
\end{small}where we have dropped the overall prefactor and there are $n-j+1$ different terms in the square brackets. 
This final expression agrees with the reconstructed result in Eq.~\eqref{reconstr-psin}.
This shows that Eq.~\eqref{eq:ampToWFnpts}  precisely matches the bulk computation when the interaction has at most one time derivative per field, although in general it will only agree up to the addition of lower-order contact terms, as discussed in Section~\ref{ssec:comments}.

An $n$-point amplitude not covered by this argument is the constant amplitude corresponding to a $\phi^n/n!$ interaction. In this case we must add analytic and log terms to the contact reconstruction formula, giving the following result for $n \geq 3$:
\be \label{eq:nPtLog}
\psi_n  =  \frac{1}{3H^4} \left[ 3\sum_{m=0}^{n-4}\frac{(n-m-4)!e_{n-m}}{k_T^{n-m-3}}-4e_3+k_T e_2-(k_T^3-3e_2 k_T+3e_3)\log(k_T/ \mu) \right].
\ee
For $n=3$ the first term in the square brackets does not contribute, giving the same result as in Eq.~\eqref{eq:bispectraLog}. The prefactor of the log corresponds to the contact wavefunction coefficient coming from the field redefinition $\phi \rightarrow \phi + \phi^{n-1}$, 
\be
\psi^{\rm local}_n = k_T^3 - 3 k_T e_2 +3 e_3 = \sum_{j=1}^n k_j^3.
\ee
This is the only expression that solves the MLT that is analytic in the energies and has the correct scaling. From this we can see that the number of contact $n$-point wavefunction coefficient is always equal to the number of contact amplitudes plus one. This is analogous to the matching between the number of CFT correlators and Lorentz-invariant amplitudes in one higher dimension \cite{Heemskerk:2009pn,Heemskerk:2010ty,Costa:2011mg}. 

 
\section{Conclusions}\label{sec:6}

The connection between physics in flat and curved spacetime plays a crucial role in cosmology. In particular, the recent advances in using general physical principles to bootstrap cosmological observables make heavy use of results and intuition from flat-space scattering amplitudes, where symmetries, locality and unitary impose very strong constraints. In cosmology, considering that correlators often have richer and more complicated singularity structures, one particularly interesting question is how one can reconstruct the answer from amplitudes in flat spacetime.

This paper contributes to the ongoing exploration of the above question. 
Fully adopting the bootstrap philosophy, we focus on the derivation of the boundary correlators of quantum fields in de Sitter space without referring to their bulk evolution.
We presented a contact reconstruction formula in Eq.~\eqref{eq:ampToWFnpts} that establishes the explicit connection between contact wavefunction coefficients and the corresponding scattering amplitudes in flat space. More precisely,  given a scalar or graviton $ n  $-particle contact amplitude from any manifestly local theory, this formula generates a corresponding cosmological correlator in de Sitter space that satisfies all relevant consistency criteria. 
To demonstrate its power, we have applied this formula to derive all possible scalar contact trispectra and we discussed how it generalises to higher-point contact correlators. This also gives us a way to count the number of independent correlators.

Several directions deserve a closer look in future investigations:
\begin{itemize}
\item It would be interesting to find a way to construct exchange wavefunction coefficients from exchange amplitudes. Such an exchange reconstruction formula would probably be much more complicated than our contact formula since it would have to reproduce the effect of many bulk time integrals, as opposed to the single time integral required for contact interactions.
A different promising approach to bootstrapping de Sitter exchange correlators is by passing through flat-space correlators. Results in this direction were recently presented in Ref.~\cite{Baumann:2021fxj} and others will appear in Ref.~\cite{toappear}.
\item In flat space we have powerful tools to constrain EFTs that admit consistent UV completions, such as positivity bounds \cite{Adams:2006sv}. It would be very exciting to understand how to rigorously derive similar bounds on cosmological backgrounds. Some progress so far has been achieved by including the breaking of boosts in flat-space amplitudes \cite{Conjecture,Baumann:2019ghk,Grall:2020tqc,Positivity2}. Hopefully a better understanding of cosmological correlators will allow us to perform the full analysis in curved spacetime.
\item It is important to better understand the role of global symmetries in cosmological correlators. In flat space, amplitudes can be strictly dictated by the spacetime and internal symmetries of a theory. 
A similar story is expected in cosmology. Since de Sitter boosts are broken by all cosmologies and all models of inflation, it is important to consistently account for their non-linear action on observables. This is well understood at the level of the Lagrangian in the EFT formalism \cite{EFTofI}. It is also understood how large diffeomorphisms constrain the soft limits of correlators, see, e.g., Refs.~\cite{Maldacena:2002vr,Hinterbichler:2012nm,Creminelli:2012ed,Assassi:2012zq,Kehagias:2012pd,Hinterbichler:2013dpa,Pajer:2017hmb,Bordin:2017ozj,Hui:2018cag,Pajer:2019jhb,Avis:2019eav}. However, it is not completely clear how to see a relation such as $  f_{\rm NL}^{{\rm eq}}\sim c_{s}^{-2} $ directly of the level of correlators (see Ref.~\cite{Berezhiani:2014tda} for some progress in this direction).
\end{itemize}
The study of cosmological correlators is still in its infancy, but exciting progress has been achieved in the past few years. We are confident that new general and insightful results are waiting to be discovered and that these will be facilitated by fertile interactions with adjacent research fields such as amplitudes and AdS/CFT.

\paragraph*{Acknowledgements} We are grateful to Ward Haddadin, Kurt Hinterbichler, Austin Joyce, Scott Melville, Guilherme Pimentel, and David Stefanyszyn for helpful discussions. 
We have been supported in part
by the research program VIDI with Project No.~680-47-535, which is (partly) financed by the
Netherlands Organisation for Scientific Research (NWO).
This work has also been partially supported by  STFC HEP consolidated grants ST/P000681/1 and ST/T000694/1.

\appendix
 
\section{The quartic wavefunction from the manifestly local test}\label{App:A}

In this appendix, we derive the four-point contact wavefunction coefficients for $  p\leq 6 $ by
solving the MLT for a general ansatz. With the help of the primary and secondary symmetric polynomials \eqref{eq:e2}-\eqref{3esp} and \eqref{primary}-\eqref{secondary}, the bootstrap rules in Section \ref{sec:2} allow us to write down a general ansatz for $\psi_4$ as
\begin{align} \label{generaltri}
&{c_{1}e_3 + { c_{2} k_Te_2}  +{c_{3} k_T^3} }
+ \(\tilde c_{1} e_3 + \tilde c_{2} k_Te_2 + \tilde c_{3} k_T^3 \) \log\( \frac{k_T}{\mu}\)\nn\\
& + \frac{1}{k_T}\(c_{4}e_4 + { c_{5} e_2^2}  +{c_{6} E_{2}} +  c_{7} S_{1}\) 
+ \frac{c_{8} e_2 e_3}{k_T^2} \nn\\
& + \frac{1}{k_T^3}\left[(c_9e_4 +  c_{10} e_2^2  + c_{11} E_{2} +  c_{12} S_{1})e_2 + c_{13} e_3^2 + c_{14} E_{3} + c_{15} S_2 + c_{16} S_3 \right] \nn\\
& +\frac{1}{k_T^4}\left(c_{17}e_4 + c_{18} e_2^2  + c_{19} E_{2} +  c_{20} S_1\right)e_3 +\frac{1}{k_T^5}\[\left(c_{21}e_4 + { c_{22} e_2^2}  +{c_{23} E_{2}} +  c_{24} S_1\right)e_4\right. \nn\\
& \left. + c_{25}e_2^4 +\( { c_{26} E_3}  +{c_{27} S_{2}} +
c_{28} S_3+ c_{29}  e_3^2 \)e_2 + \(c_{30}  E_2 + c_{31}  S_1\)e_2^2 + c_{32} S_1^2+  c_{32} E_2^2 + c_{33} S_4 \] \nn\\
&+\frac{1}{k_T^6} \left[(c_{34}e_4 +  c_{35} e_2^2  + c_{36} E_{2} +  c_{37} S_{1})e_2e_3  + \(c_{38} E_{3} + c_{39} S_2+ c_{40} S_3\) e_3+ c_{41} e_3^3 \right] ,
\end{align}
with 44 free parameters $c_i$ and $\tilde{c}_i$. The terms are organised by the degree of their total-energy pole.
For quartic interactions with the maximum number of derivatives $q_{\rm max}$, the  $\psi_4$ ansatz should contain all possible terms up to $1/k_T^p$, with $p=q_{\rm max}+1$.
Meanwhile, the residue of the leading $k_T$ pole is fixed by the corresponding manifestly local amplitudes $A_4$, as in Eq.~\eqref{eq:leading-pole}.
There must be at least one power of $e_4$ in the numerator of the  $1/k_T^p$ term, as required by the MLT. For the subleading $k_T$ poles the constraint from locality is more subtle and all terms in the ansatz \eqref{generaltri} should be included.

To illustrate how the coefficients on the subleading $k_T$ poles are fixed by a given flat-space amplitude, here we explicitly solve the MLT case by case from $p=0$ to $6$. 
Using this approach, we derive  all  possible wavefunction coefficients for contact interactions with up to five derivatives, reproducing the results from the contact reconstruction formula in Section \ref{sec:trispectrum}.
\begin{itemize}
\item $p=0$. Let us first consider the case without $k_T$ poles, where
the bootstrap ansatz can be simply written as
\be \label{psip0}
\psi^{\rm ansatz}_4 
= 
{c_{1}e_3 + { c_{2} k_Te_2}  +{c_{3} k_T^3} } .
\ee
One could include terms proportional to $ \log(k_{T})  $, but the MLT demands that they all vanish.
Applying the MLT gives the constraints
\be
c_{1} +c_{2} =0, ~~~~~~~~ c_{2} +3c_{3}=0 .
\ee
Therefore at this order there is only one allowed shape,
\be
\mathcal{S}^{\rm local}_4 
\equiv 
3e_3 - { 3 k_Te_2}  +{k_T^3}  ,
\ee
which corresponds to the local non-Gaussianity that arises  from the field redefinition $\phi \rightarrow \phi + \phi^3$.

\item $p=1$. 
At this order the new bootstrap ansatz to add to Eq.~\eqref{psip0} is given by
\be
 \(\tilde c_{1} e_3 + \tilde c_{2} k_Te_2 + \tilde c_{3} k_T^3 \) \log\( \frac{k_T}{\mu}\) + \frac{1}{k_T}\(c_{4}e_4 + { c_{5} e_2^2}  +{c_{6} E_{2}} +  c_{7} S_{4}\) .
\ee
Solving the MLT yields the following constraints:
\be
&&c_{5}= c_{6} =c_{7} =0, ~~~~~~~~  \tilde c_{1} +\tilde c_{2} =0, ~~~~~~~~ \tilde c_{2} +3\tilde c_{3}=0 ,\\
&&c_{4} +  \tilde c_{1} = 0 , ~~~~~~~~ c_{1} +c_{2} + \tilde c_{2} =0 , ~~~~~~~~ c_{2} +3c_{3}+ \tilde c_{3}=0 .
\ee
The final result can be written in terms of two independent shapes,
\be
\psi^{p\leq 1}_4 
=  \tilde c_{3} \mathcal{S}^{\rm p=1}_4 
+c_{3} \mathcal{S}^{\rm local}_4 ,
\ee
where we define the new shape arising at this order as
\be
\mathcal{S}^{\rm p=1}_4 \equiv -3 \frac{e_4}{k_T} +4 e_3 - k_T e_2
+ \( k_T^3 - 3 k_T e_2 +3  e_3 \) \log\( \frac{k_T}{\mu}\) ,
\ee
which is generated by the $\phi^4$ interaction in de Sitter space.

\item $p=2$. There is no four-particle amplitude with one derivative, thus no wavefunction coefficient is expected at this order. Another way to see this fact is that
in the ansatz the
only possible term for the leading $k_T$ pole is $c_8{e_2 e_3}/{k_T^2} $, while
the
MLT requires $c_{8}=0$, thus no new  shape arises at this order.

\item $p=3$. In the ansatz the leading $k_T$ pole of this order is
\be
c_9\frac{e_4e_2}{k_T^3} ,
\ee
while the subleading terms in the ansatz are given by the first two lines in Eq.~\eqref{generaltri}.
After applying the MLT and solving for the relations of $c_i$, the wavefunction coefficient reduces to three independent parts,
\be
\psi^{p\leq3}_4 
= c_{9} \mathcal{S}^{\rm p=3}_4 + \tilde c_{3} \mathcal{S}^{\rm p=1}_4 
+c_{3} \mathcal{S}^{\rm local}_4 ,
 \ee
where the new shape at this order is given by
\be \label{psi43}
\mathcal{S}^{\rm p=3}_4 \equiv \frac{e_4e_2}{k_T^3}+\frac{e_2e_3}{2k_T^2} +\frac{e_2^2-e_4}{2k_T}-e_3 ,
\ee
which agrees with Eq.~\eqref{psi43r} from the contact reconstruction formula.

\item $p=4$. At this order we need to include the term
\be
c_{17}\frac{e_4e_3}{k_T^4} ,
\ee
while the rest of the ansatz is given by the first three lines of Eq.~\eqref{generaltri}.
After applying the MLT, the final result can be written as linear combinations of four independent parts
\be
\psi^{p\leq4}_4 = c_{17} \mathcal{S}^{\rm p=4}_4 +
c_{9} \mathcal{S}^{\rm p=3}_4 + \tilde c_{3} \mathcal{S}^{\rm p=1}_4 
+c_{3} \mathcal{S}^{\rm local}_4 ,
\ee
where the new shape arising at this order is given by
\be \label{psi44}
\mathcal{S}^{\rm p=4}_4 \equiv \frac{e_4e_3}{k_T^4}+\frac{e_3^2}{3k_T^3} +\frac{e_3e_2}{3k_T^2}+\frac{e_2^2-e_4}{3k_T} -\frac{2}{3}e_3 .
\ee
In terms of the results from the contact reconstruction formula, Eqs.~\eqref{psi43r} and \eqref{psi44r}, it can be expressed as
\be
\mathcal{S}^{\rm p=4}_4 = \mathcal{S}^{(4)}_4 + \frac{2}{3} \mathcal{S}^{(3)}_4 .
\ee

\item $p=5$. 
Here the leading $k_T$-pole terms contain four different contributions 
\be
\frac{\left(c_{21}e_4 + { c_{22} e_2^2}  +{c_{23} E_{2}} +  c_{24} S_1\right)e_4}{k_T^5} ,
\ee
while the complete ansatz at this order corresponds to the first four lines of Eq.~\eqref{generaltri}.
The constraints from the MLT reduce this ansatz to the following 8 independent parts
\be
\psi^{p\leq5}_4 
&=& 
c_{21} \mathcal{S}^{\rm  p=5|a}_4 + c_{22} \mathcal{S}^{\rm  p=5|b}_4 + c_{23} \mathcal{S}^{\rm  p=5|c}_4 + c_{24} \mathcal{S}^{\rm  p=5|d}_4\nn\\
&&+
c_{13} \mathcal{S}^{\rm p=4}_4 +
c_{8} \mathcal{S}^{\rm p=3}_4 + \tilde c_{3} \mathcal{S}^{\rm p=1}_4 
+c_{3} \mathcal{S}^{\rm local}_4 .
\ee
The four new contributions are related to the four $k_T$-pole terms of degree $p=5$,
\begin{align}
&\mathcal{S}^{\rm  p=5|a}_4 \equiv  \frac{e_4^2}{k_T^5} , \label{psi45a}\\
&\mathcal{S}^{\rm  p=5|b}_4 \equiv  \frac{e_2^2e_4}{k_T^5} + \frac{e_2^2e_3}{4k_T^4}-\frac{e_2e_4}{2k_T^3} + \frac{e_2^3}{12k_T^3} + \frac{e_2^2}{4k_T}-\frac{e_3}{2} , \label{psi45b}\\
&\mathcal{S}^{\rm  p=5|c}_4 \equiv  \frac{E_{2}e_4}{k_T^5} + \frac{E_{2}e_3}{4k_T^4}+ \frac{E_{2}e_2}{12k_T^3} + \frac{E_{2}}{12k_T} ,  \label{psi45c}\\
&\mathcal{S}^{\rm  p=5|d}_4 \equiv   \frac{S_1 e_4}{k_T^5} + \frac{S_1 e_3}{4k_T^4}- \frac{2e_4e_2+e_2 S_1+S_3}{12k_T^3} + \frac{e_4-2e_2^2 + S_1}{12k_T} + \frac{e_3}{4} , \label{psi45d}
\end{align}
which are related to the reconstructed results \eqref{psi45ar}-\eqref{psi45dr} by the following linear combinations with lower-order wavefunction coefficients:
\be
&&\mathcal{S}^{\rm p=5|a}_4 = \mathcal{S}^{(5a)}_4  ,~~~~ \mathcal{S}^{\rm  p=5|b}_4 = \mathcal{S}^{(5b)}_4  - \frac{3}{4}\mathcal{S}_4^{(4)} +\frac{1}{2}\mathcal{S}^{(3)} ,~~~~ \nn\\
&&\mathcal{S}^{\rm  p=5|c}_4 = \mathcal{S}^{(5c)}_4  ,~~~~ \mathcal{S}^{\rm  p=5|d}_4 = \mathcal{S}^{(5d)}_4  - \frac{1}{3}\mathcal{S}_4^{(3)} .
\ee

\item $p=6$. Here the full ansatz is given by  Eq.~\eqref{generaltri}, but there is only one leading $k_T$-pole term,
\be
c_{34}\frac{e_4e_2e_3}{k_T^6} .
\ee
The final wavefunction coefficient has one additional component,
\be
\psi^{p\leq6}_4 
&=& 
c_{34} \mathcal{S}^{\rm  p=6}_4+
c_{21} \mathcal{S}^{\rm  p=5|a}_4 + c_{22} \mathcal{S}^{\rm  p=5|b}_4 + c_{23} \mathcal{S}^{\rm  p=5|c}_4 + c_{24} \mathcal{S}^{\rm  p=5|d}_4\nn\\
&&+
c_{13} \mathcal{S}^{\rm p=4}_4 +
c_{8} \mathcal{S}^{\rm p=3}_4 + \tilde c_{3} \mathcal{S}^{\rm p=1}_4 
+c_{3} \mathcal{S}^{\rm local}_4  ,
\ee
with 
\be
\mathcal{S}_4^{\rm  p=6} \equiv  \frac{e_4e_2e_3}{k_T^6} + \frac{e_2e_3^2}{5k_T^5}+ \frac{e_2^2e_3}{10k_T^4}
+ \frac{e_2^3+2e_3^2}{30k_T^3}  + \frac{e_2e_3}{6k_T^2}
+ \frac{4e_2^2}{15k_T} -\frac{e_4}{6k_T} -\frac{8}{15}e_3 .
\ee
This new shape  agrees with Eq.~\eqref{psi46r} from the contact reconstruction formula when we take the following linear combination with lower-order contributions:
\be
\mathcal{S}^{\rm  p=6}_4 =  \mathcal{S}^{(6)}_4 -\frac{4}{5} \mathcal{S}^{(5a)}_4  +\frac{2}{5} \mathcal{S}^{(5b)}_4 - \frac{1}{10}  \mathcal{S}^{(4)}_4 +\frac{8}{15}  \mathcal{S}_4^{(3)} .
\ee

\end{itemize}


\bibliographystyle{utphys}
\bibliography{refs}

\end{document}